\documentclass[camera,letterpaper,nomarginnotes,nonarrowgutter]{jpaper}
\usepackage{xspace}
\usepackage{tikz}
\usepackage{xcolor}
\usepackage{soul}
\usepackage{multicol}
\usepackage{multirow}
\usepackage{booktabs}
\usepackage{fancyhdr}
\usepackage{pdfpages}
\usepackage[nomessages]{fp}
\usepackage{tocbibind}
\usepackage{clipboard}

\usepackage{dblfloatfix}
\usepackage{mathptmx} 
\usepackage[normalem]{ulem}
\usepackage{url}
\usepackage{marginnote}
\usepackage{enumitem}
\usepackage{ifthen}
\usepackage{caption}
\usepackage{listings}
\usepackage{setspace}
\usepackage{float}
\usepackage[us,12hr]{datetime}
\usepackage[textwidth=1cm, colorinlistoftodos,prependcaption,textsize=small]{todonotes}
\usepackage{xargs}
\usepackage{titletoc}
\usepackage[all]{nowidow}
\usepackage[framemethod=tikz]{mdframed}
\usepackage[most]{tcolorbox}
\tcbuselibrary{breakable}
\tcbuselibrary{hooks}%

\usepackage{mathtools}

\usepackage{courier}
\usepackage[italic]{mathastext}

\usepackage{siunitx}
\usepackage{arydshln} %

\usepackage{pifont}
\usepackage{footmisc}
\usepackage{hhline} %
\usepackage[compact]{titlesec} %
\usepackage{footnote}

\usepackage{balance}
\usepackage[utf8]{inputenc} %
\usepackage[nocompress]{cite}
\usepackage{multirow}
\usepackage{hhline}
\usepackage{textcomp}
\usepackage{subcaption} 
\usepackage{fancyhdr}
\usepackage[normalem]{ulem}
\usepackage{array}
\usepackage{ragged2e}
\usepackage{array}
\usepackage{pifont}%
\usepackage{amsmath}
\usepackage[linesnumbered,ruled]{algorithm2e}
\usepackage{algpseudocode}

\usepackage[colorlinks,citecolor=black,urlcolor=blue,bookmarks=false,hypertexnames=true,linkcolor=black,anchorcolor=black,filecolor=black]{hyperref}
\usepackage[resetlabels]{multibib}

\definecolor{denim}{rgb}{0.08, 0.38, 0.74}
\definecolor{darkolivegreen}{rgb}{0.33, 0.42, 0.18}
\definecolor{dgreen}{rgb}{0.00, 0.75, 0.00}
\definecolor{darkpink}{rgb}{0.88, 0.28, 0.54}
\definecolor{forestgreen}{rgb}{0.0, 0.27, 0.13}
\definecolor{amber}{rgb}{1.0, 0.49, 0.0}
\definecolor{lightyellow}{rgb}{0.980, 0.956, 0.623}
\definecolor{lightblue}{rgb}{0.980, 0.956, 0.623}
\definecolor{darkamber}{rgb}{0.5, 0.19, 0.0}
\definecolor{dkgreen}{rgb}{0,0.6,0}
\definecolor{gray}{rgb}{0.5,0.5,0.5}
\definecolor{mauve}{rgb}{0.58,0,0.82}
\definecolor{lightmauve}{rgb}{0.68,0.4,0.92}
\definecolor{chocolate}{rgb}{0.48, 0.25, 0.0}
\definecolor{dollarbill}{rgb}{0.52,0.73,0.4}
\definecolor{dkdkgreen}{rgb}{0,0.45,0}
\definecolor{gfored}{rgb}{0.580, 0.050, 0.211}
\definecolor{darkwarmgray}{rgb}{0.15, 0.050, 0.05}
\definecolor{ups-truck}{rgb}{0.53, 0.28, 0.21}

\newcommand\rev[1]{{\color{black}{#1}}}

\newcommand\ltitle{AirLift: A Fast and Comprehensive Technique\\for Remapping Alignments between Reference Genomes\xspace}

\newcommand{\release}{\href{https://github.com/CMU-SAFARI/AirLift}{https://github.com/CMU-SAFARI/AirLift}\xspace}

\newcommand\ovmaxseed{$27.4\times$\xspace}

\newcommand{\figlimitations}{1}

\let\oldmarginnote\marginnote

\renewcommand{\marginnote}[2][rectangle,draw,fill=blue!40,rounded corners]{%
    \oldmarginnote{%
    \tikz \node at (0,0) [#1]{#2};}%
}

\titlespacing*{\section}{0pt}{2pt plus 0.5pt minus 0.5pt}{0pt}
\titlespacing*{\subsection}{0pt}{2pt plus 0.5pt minus 0.5pt}{0pt}
\titlespacing*{\subsubsection}{0pt}{2pt plus 0.5pt minus 0.5pt}{0pt}

\makeatletter
\g@addto@macro{\normalsize}{%
  \setlength{\abovedisplayskip}{2pt plus 1pt minus 1pt}
  \setlength{\belowdisplayskip}{2pt plus 1pt minus 1pt}
  \setlength{\abovedisplayshortskip}{0pt}
  \setlength{\belowdisplayshortskip}{0pt}
  \setlength{\intextsep}{2pt plus 1pt minus 1pt}
  \setlength{\textfloatsep}{3pt plus 1pt minus 1pt}
  \setlength{\dbltextfloatsep}{3pt plus 1pt minus 1pt}
  \setlength{\skip\footins}{4pt plus 1pt minus 1pt}}
\makeatother

\renewcommand\thefootnote{{\arabic{footnote}}}

\newcommand*\circled[1]{\tikz[baseline=(char.base)]{
            \node[shape=circle,draw,inner sep=2pt] (char) {\footnotesize \textbf{#1}};}}

\newcommand{\cmark}{\ding{51}}%
\newcommand{\xmark}{\ding{55}}%

\expandafter\def\expandafter\UrlBreaks\expandafter{\UrlBreaks
  \do\a\do\b\do\c\do\d\do\e\do\f\do\g\do\h\do\i\do\j
  \do\k\do\l\do\m\do\n\do\o\do\p\do\q\do\r\do\s\do\t
  \do\u\do\v\do\w\do\x\do\y\do\z\do\A\do\B\do\C\do\D
  \do\E\do\F\do\G\do\H\do\I\do\J\do\K\do\L\do\M\do\N
  \do\O\do\P\do\Q\do\R\do\S\do\T\do\U\do\V\do\W\do\X
  \do\Y\do\Z}

\newcommand{\squishlist}{
 \begin{list}{$\circ$}
  { \setlength{\itemsep}{0pt}
     \setlength{\parsep}{0pt}
     \setlength{\topsep}{0pt}
     \setlength{\partopsep}{0pt}
     \setlength{\leftmargin}{1em}
     \setlength{\labelwidth}{1em}
     \setlength{\labelsep}{0.5em} } }

\newcommand{\squishsublist}{
\begin{list}{$\rightarrow$}
 { \setlength{\itemsep}{0pt}
    \setlength{\parsep}{0pt}
    \setlength{\topsep}{-10em}
    \setlength{\partopsep}{-3pt}
    \setlength{\leftmargin}{1em}
    \setlength{\labelwidth}{1em}
    \setlength{\labelsep}{0.5em} } }

\let\oldthebibliography\thebibliography
\renewcommand\thebibliography[1]{
  \oldthebibliography{#1}
  \setlength{\parskip}{0pt}
  \setlength{\itemsep}{0pt plus 0.3ex}
}

\newcommand{\squishend}{
  \end{list}  }
  
\fancypagestyle{firstpage}{

  \pagenumbering{arabic}
}
 
\titlespacing\section{2pt}{3pt plus 1pt minus 1pt}{2pt plus 1pt minus 1pt}
\titlespacing\subsection{2pt}{3pt plus 1pt minus 1pt}{2pt plus 1pt minus 1pt}
\titlespacing\subsubsection{2pt}{3pt plus 1pt minus 1pt}{2pt plus 1pt minus 1pt}

\newcommandx{\changev}[2][1=]{\todo[linecolor=blue,backgroundcolor=blue!25,bordercolor=blue,#1,size=\scriptsize]{#2}}

\let\oldmarginnote\marginnote

\renewcommand{\marginnote}[2][rectangle,draw,fill=blue!40,rounded corners]{%
        \oldmarginnote{%
        \tikz \node at (0,0) [#1]{#2};}%
        }
        
\newcommand{\boxbegin} {
	\begin{tcolorbox}[enhanced, frame hidden, colback=gray!50, breakable]
}

\newcommand{\boxend} {
	\end{tcolorbox}
}

\newcommand{\yboxbegin} {
	\begin{tcolorbox}[breakable, enhanced, frame hidden, colback=yellow!50]
}

\newcommand{\yboxend} {
	\end{tcolorbox}
}

\mdfdefinestyle{graybox}{
    splittopskip=0,%
    splitbottomskip=0,%
    frametitleaboveskip=0,
    frametitlebelowskip=0,
    skipabove=0,%
    skipbelow=0,%
    leftmargin=0,%
    rightmargin=0,%
    innertopmargin=0mm,%
    innerbottommargin=0mm,%
    roundcorner=1mm,%
    backgroundcolor=lightblue,
    hidealllines=true}
    
\mdfdefinestyle{graybox2}{
    splittopskip=0,%
    splitbottomskip=0,%
    frametitleaboveskip=0,
    frametitlebelowskip=0,
    skipabove=0,%
    skipbelow=0,%
    leftmargin=0,%
    rightmargin=0,%
    innertopmargin=2mm,%
    innerbottommargin=2mm,%
    roundcorner=2mm,%
    backgroundcolor=lightblue,
    hidealllines=true}

\newcommand{\bboxbegin}{
    \begin{mdframed}[style=graybox]
}

\newcommand{\bboxend}{
    \end{mdframed}
}

\newcommand{\yyboxbegin}{
    \begin{mdframed}[style=graybox2]
}

\newcommand{\yyboxend}{
    \end{mdframed}
}

\newcolumntype{P}[1]{>{\centering\let\newline\\\arraybackslash}p{#1}}
\newcolumntype{R}[1]{>{\raggedright\let\newline\\\arraybackslash}p{#1}}

\definecolor{darkwarmgray}{rgb}{0.15, 0.050, 0.05}
\definecolor{dollarbill}{rgb}{0.52, 0.73, 0.4}
\newcommand{\jkt}[1]{{\color{black}#1}}

\newcommand{\pair}[2]{\texttt{#1$\rightarrow$#2}}

\newcommand{\blackcircled}[1]{\tikz[baseline=(char.base)]{\node[shape=circle,inner sep=1pt,fill=darkwarmgray, text=white] (char) {\footnotesize \textbf{#1}};}}

\newif\ifcameraready
\camerareadyfalse

\newcommand{\squeezeme}{ \setlength{\itemsep}{0pt}
     \setlength{\parsep}{3pt}
     \setlength{\topsep}{3pt}
     \setlength{\partopsep}{0pt}
     \setlength{\leftmargin}{1.5em}
     \setlength{\labelwidth}{1em}
     \setlength{\labelsep}{0.5em} }

\let\oldtableofcontents\tableofcontents%
\renewcommand\tableofcontents{
  \oldtableofcontents%
  \clearpage
}

\newcommand\blfootnote[1]{%
  \begingroup
  \renewcommand\thefootnote{}\footnote{#1}%
  \addtocounter{footnote}{-1}%
  \endgroup
}

\makeindex

\newcommand{\affilETH}[0]{\small {$^1$}}
\newcommand{\affilCMU}[0]{\small {$^2$}}
\newcommand{\affilSFU}[0]{\small {$^3$}}
\newcommand{\affilBilkent}[0]{\small {$^4$}}
\newcommand{\equalcont}[0]{\small {$^{\dagger}$}}
\title{\ltitle} 
\author{\vspace{-17pt}\\%
\fontsize{11}{12}\selectfont%
{Jeremie S. Kim\affilETH{}$^{,}$\equalcont{}}\quad%
{Can Firtina\affilETH{}$^{,}$\equalcont{}}\quad%
{Meryem Banu Cavlak\affilETH{}}\quad%
{Damla Senol Cali\affilCMU{}}\quad%
\vspace{-1pt}\\%
\fontsize{11}{12}\selectfont%
{Nastaran Hajinazar\affilETH{}$^{,}$\affilSFU{}}\quad%
{Mohammed Alser\affilETH{}}\quad%
{Can Alkan\affilBilkent{}}\quad%
{Onur Mutlu\affilETH{}$^{,}$\affilCMU{}$^{,}$\affilBilkent{}}%
\vspace{-1pt}\\%
{\fontsize{10}{11}\selectfont
\affilETH\emph{ETH Zurich}%
\qquad\quad%
\affilCMU\emph{Carnegie Mellon University}%
\qquad\quad%
\affilSFU\emph{Simon Fraser University}%
\qquad\quad%
\affilBilkent\emph{Bilkent University}%
}
\vspace{-16pt}\vspace{0.3em}}

\newcites{supp}{Supplementary References}

\begin{document}
\maketitle
\thispagestyle{plain}
\pagestyle{plain}
\setstretch{0.86}

\begin{abstract}
AirLift is the first read remapping tool that enables users to quickly and
comprehensively map a read set, that had been previously mapped to one
reference genome, to another similar reference. Users can then quickly run
a downstream analysis of read sets for each latest reference release.  Compared
to the state-of-the-art method for remapping reads (i.e., full mapping),
AirLift reduces the overall execution time to remap read sets between two
reference genome versions by up to \ovmaxseed.  We validate our remapping results with GATK and find that AirLift provides high accuracy in identifying ground truth SNP/INDEL variants.

\noindent\textbf{Code Availability.} AirLift source code and readme describing how to reproduce our results are available at \release.
\end{abstract}

\blfootnote{\equalcont\emph{Equal Contributor}}
\section{Introduction} \label{sec:intro}

Reference genomes are inaccurate and do not perfectly represent the average
healthy individual of a species for a variety of
reasons~\cite{mallick2016simons, sherman2019assembly}. 
First, reference genomes are constructed using imperfect sequencing
technologies that result in error-prone reads~\cite{ma2019analysis, cali2019nanopore}. Second,
the sequenced reads of an individual (i.e., \emph{read set}) are assembled into
a reference genome using imperfect assembly tools~\cite{alkan2011limitations,
steinberg2017building}. As genome sequencing technology and assembly algorithms
improve, and as more sequenced samples become available, researchers are able
to incrementally assemble more accurate reference genomes. As an example, the
Genome Reference Consortium (GRC) releases minor updates to the human reference
genome every three months and major updates every few
years~\cite{RefSeqCuration, GRC_patches}. \rev{Recently,} significant advances
have resulted in a novel full telomere to telomere
reference~\cite{miga2020telomere}. These updates are \emph{critical} to the
accuracy of the reference genome as they enable the latest reference genome to
provide the most accurate and complete representation of the reference's
respective population. Therefore, a read set should be mapped to the latest and
most relevant reference genome to obtain the most accurate downstream genome
analysis results~\cite{guo2017improvements}. 

Currently, the best way to adapt an existing genomic study (i.e., read sets
from many samples) to a new reference genome is to re-run the \emph{entire}
analysis pipeline using the new reference genome.
For example, 
after completing the 1000 Genomes Project using the human reference genome build 37 (GRCh37)~\cite{10002015global}, 
each read set from this project was mapped again to the next version of the human
reference genome (GRCh38)~\cite{zheng2017alignment}. Unfortunately, this
approach is \emph{computationally very expensive} and does not scale to large
genomic studies that include a large number of individuals for three key
reasons. First, mapping even a \emph{single} read set is computationally
expensive~\cite{Ruffalo2011, canzar2015short} (e.g., 75 hours for aligning
300,000,000 short reads, which provides 30$\times$ coverage of the human
genome~\cite{miga2020telomere}) as it heavily relies on a computationally-costly alignment
algorithm~\cite{alser2020technology, alser2020accelerating, firtina_blend_2021}. Second, the
number of available read sets doubles approximately every 8
months~\cite{sequencing_rate1, sequencing_rate2}, and the rate of growth will
continue to increase as sequencing technologies continue to become more cost
effective and higher throughput~\cite{cali2019nanopore}.  Third,
researchers are beginning to use highly specific reference genomes that better
represent diverse populations and ethnic groups~\cite{al2013genome,
xu2014genome, ahn2009first, sherman2019assembly, wang2008diploid,
schuster2010complete, huang2015genetic, shukla2019hg19kindel}. This may
result in the need to map each read set to \emph{multiple} reference genomes
that represent various populations within the same species in order to
correctly identify the genome donor's genetic variations (i.e., differences
from the most relevant reference genome). 

\begin{figure}[htb]
    \centering
	\includegraphics[width=\linewidth]{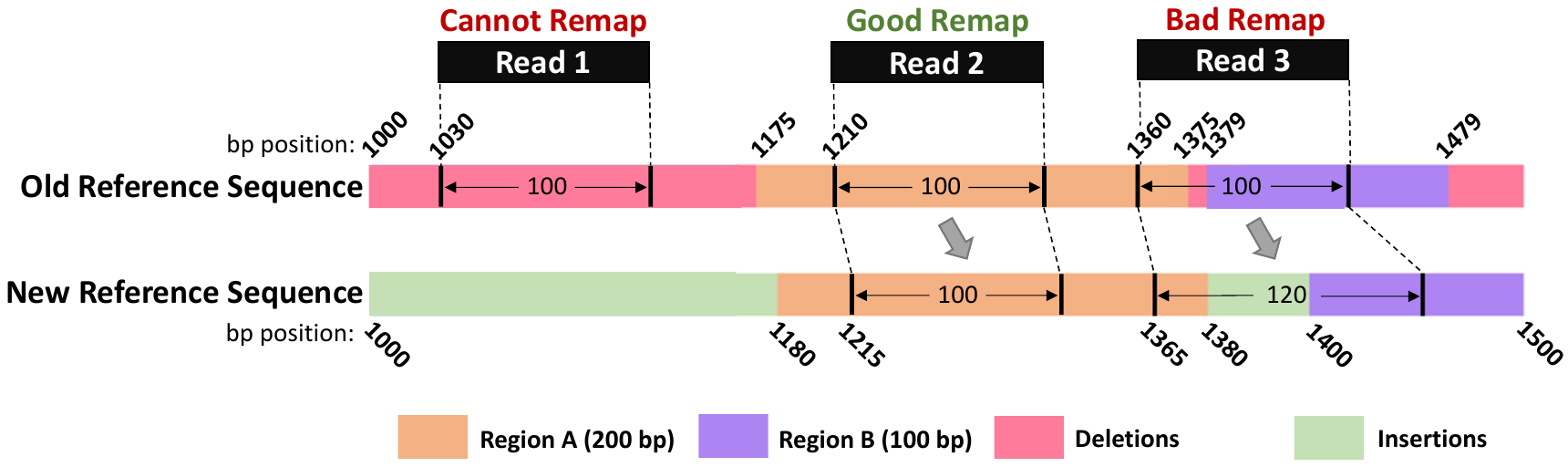} 
	\caption{Limitations of Existing Remapping Tools. Existing remapping tools correctly remap reads that mapped \emph{completely within} a region indicated by the chain file (e.g., Read 2). However, these tools 1) cannot remap reads that mapped within a region in the old reference that does not appear in the new reference (e.g., Read 1) and 2) may incorrectly remap reads that align to multiple constant regions in the old reference (e.g., Read 3).} 
	\label{fig:chain_file_downsides} 
\end{figure}

To reduce the large overhead of \emph{fully mapping} a read set to a new
reference genome, several existing tools~\cite{ucscliftover, crossmap,
segmentliftover, gao2018segment_liftover, zhao2013crossmap, ncbi_genome_remap,
galaxy_remap, pyliftover_remap, mun_leviosam_2021, chen_improved_2022} can be used to quickly \emph{remap} the reads
(i.e., update a read's alignment location from the original (old) reference to
another (new) reference). 
In the remainder of this paper, we collectively refer to such methods as
\textit{remapping tools}. At a high level, state-of-the-art remapping tools
rely on \emph{chain files} (described in Supplementary
Section S2), which identify and list \emph{constant
regions}, i.e., genome sequences that appear in both old and new references
(e.g., regions A and B in Figure~\ref{fig:chain_file_downsides}) and
their positional offsets into each reference genome. A remapping tool uses a
chain file to identify reads whose original mapping locations in the old
reference is sufficiently contained within constant regions and quickly
updates the alignment location of each read according to how the location of
the constant region containing it changes between the old and new references.
For example, Read 2 in Figure~\ref{fig:chain_file_downsides} can be quickly
remapped by shifting its location by 5 base pairs from the old reference to the
new reference (these tools are described in more detail in
Supplementary Section S1.)

Unfortunately, many of these remapping tools 1) are not \emph{comprehensive} in
remapping a read set, meaning that they \emph{cannot} remap a significant
proportion of reads due to the limitations of using a chain file (e.g., a chain
file only contains information about genome sequences that appear exactly the
same between two references and their positional offsets in each reference),
2) are not \emph{accurate}, meaning that some remapped reads do \emph{not}
align to the sequence they are remapped to in the new reference genome within
the acceptable error rate, and 3) result in output on which downstream analysis
cannot be performed (i.e., do not provide an end-to-end BAM-to-BAM\footnote{A
BAM file is the binary version of a SAM file. A SAM file is a tab-delimited
text file that contains sequence alignment data~\cite{bamspecs}.} remapping
solution).  We identify two key limitations that we illustrate in
Figure~\ref{fig:chain_file_downsides}.  First, since each \emph{deleted region}
(i.e., a region that does not appear in the new reference) does not have a
corresponding region in the new reference, chain files \emph{cannot} provide
information on how to remap reads that had originally mapped to a deleted
region. This is because, by definition, a deleted region has no similar regions
in the new reference. For example, Read 1 in
Figure~\ref{fig:chain_file_downsides} maps to a deleted region in the old
reference and, therefore, cannot be remapped to the new reference to any extent.
Second, state-of-the-art remapping tools \emph{only} consider the degree of
similarity between a read and the constant regions (from the chain file) in the
old reference, without considering the changes in the new reference when
remapping the read to the new reference.  Therefore, remapping can result in a
poor degree of similarity between the read and the new reference. As an
example, Read 3 in Figure~\ref{fig:chain_file_downsides} maps to the old
reference with high similarity (i.e., 4 deletions between base pairs 1375 and
1379; $<5\%$ error rate), so it is remapped to the new reference at a location
corresponding to the read's original mapping in the old reference. This
remapping does not account for differences that appear in the new reference
(e.g., 20 insertions between base pairs 1380 and 1400) and result\rev{s} in a high
error rate (i.e., $>5\%$). 

Due to these limitations, existing remapping tools are unable to
comprehensively remap a read set from one reference to another. We observe that
state-of-the-art remapping tools miss at least 7\% of gene annotations when
remapping reads from an older human reference genome (hg16) to its latest
version (GRCh38), as shown in Supplementary Table S1 and
Supplementary Figure S1. These limitations require
researchers and practitioners to re-run the \emph{full} genome analysis
pipeline for each read set on an updated reference genome for a comprehensive
study.

Our \textbf{goal} is to provide the first read remapping technique
across (reference) genomes 1) that \textit{substantially} reduces the time
to remap a read set from an old (i.e., previously mapped to) reference genome
to a new reference genome, 2) that is \emph{comprehensive} in remapping a read
set, i.e., attempts to remap \emph{all} reads in a read set, 3) provides
\emph{accurate} remapping results, i.e., provides alignments with error rates
below a specified acceptable error rate, and 4) provides an end-to-end
BAM-to-BAM remapping solution on which downstream analysis can be immediately
performed. To this end, we propose \emph{AirLift}, the first methodology and
tool that leverages the similarity between two reference genomes to satisfy our
goal. Specifically, AirLift greatly reduces the time to perform end-to-end
BAM-to-BAM remapping on a read set from one reference genome to another while
maintaining high accuracy and comprehensiveness that is comparable to
\emph{fully mapping} the read set to the new reference.

We evaluate AirLift and demonstrate that AirLift satisfies the four design
goals of an effective remapping tool by comparing it against
state-of-the-art remapping tools and the previous best method of \emph{fully
mapping} a read set to a new reference with BWA-MEM~\cite{li2013aligning}
across various versions of the human, C. elegans, and yeast references
(summarized in Table~\ref{tab:prior_dram_works}). We demonstrate that AirLift
can identify SNPs and Indels with precision and recall similar to full mapping
(via GATK HaplotypeCaller~\cite{McKenna2010}) while providing $2.76\times$
to $27.4\times$ speedup over \emph{fully mapping} a read set to the new
reference genome.

\begin{table}[!ht]
\scriptsize
\begin{center}
\resizebox{\linewidth}{!}{
\begin{tabular}{ r|ccccc }
      \textbf{Mechanism}
	
	& \textbf{Fast}
	& \textbf{Comprehensive} 
	& \textbf{Accurate}
    & \textbf{BAM-to-BAM}
	& \textbf{Memory Usage} \\ 
\hline 
CrossMap~\cite{zhao2013crossmap}
	
	& \cmark  
	& \xmark 
	& \xmark  
    & \xmark 
    & low \\ 
\hline
LiftOver~\cite{ucscliftover} 
	
	& \cmark  
	& \xmark
	& \xmark  
    & \xmark 
    & low \\ 
\hline
Full Mapping 
	
	& \multirow{2}{*}{\xmark} 
	& \multirow{2}{*}{\cmark}
	& \multirow{2}{*}{\cmark}  
    & \multirow{2}{*}{\cmark} 
    & \multirow{2}{*}{high} \\ 
(BWA-MEM~\cite{li2013aligning}) 
	
	&   
	& 
	&   
    &  
    & \\  
\hline
\textbf{AirLift} 
	
    & \cmark 
	& \cmark 
	& \cmark  
    & \cmark 
	& high  \\ 
\hline
\end{tabular}}
\caption{AirLift vs. existing state-of-the-art remapping tools.}
\label{tab:prior_dram_works}
\end{center}
\end{table}

\section{AirLift} 
\label{sec:mechanism}

In order to accurately and comprehensively remap a read set, AirLift 1)
categorizes and labels each region (i.e., a contiguous sequence within a
genome) in the old reference genome, depending on its degree of similarity to
the most similar region in the new reference and 2) remaps each read from the
old reference to the new reference according to the label of the region in the
old reference that the read had been originally mapped to. 

For each pair of references that AirLift remaps reads between, we must
first construct an \emph{AirLift Index}, i.e., a set of lookup tables (LUTs),
in a one-time preprocessing step. AirLift queries the AirLift Index with a read
and its original mapping location in the old reference (from the BAM file) to
efficiently identify the region and the label of the region that the read
mapped to in the old reference. This information is then used to identify
potential mapping locations of the read in the new reference (based on regions
in the new reference that are similar to the region that the read mapped to in
the old reference). 

We next define these regions, show how to generate the \emph{AirLift
Index}, and then explain how to use the \emph{AirLift Index} to quickly
remap a read set with high genome coverage.  

\subsection{Reference Genome Regions} 
\label{subsec:genome_regions}

\begin{figure}[ht]
    \centering
	\includegraphics[width=\linewidth]{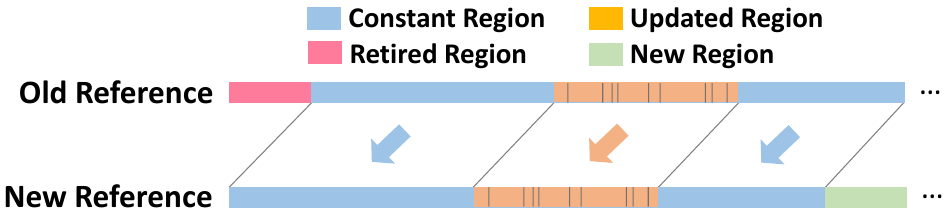}
	\caption{An example pair of reference genomes (old and new) with regions labeled (as constant, updated, retired, and new regions) and associated with each other according to their degrees of similarity. Regions that are associated with (i.e., similar to) each other are indicated with an arrow. Example differences across associated updated regions are shown with black vertical bars.}
	\label{fig:genome_regions} 
\end{figure}

We identify four categories of regions that fully describe the relationship
between two reference genomes, old and new (shown in
Figure~\ref{fig:genome_regions}):
\begin{enumerate}
\squeezeme
\item A \textit{constant region} is a region of the genome that is exactly the same in both old and new reference genomes (colored in blue). The start and end positions of a constant region are not necessarily the same in the old and new reference genomes. 
\item An \textit{updated region} is a region in the old reference genome that maps to at least one region in the new reference genome within a reasonable error rate, i.e., differences from the old reference (colored in orange with some differences marked with black bars). 
\item A \textit{retired region} is a region in the old reference genome that does \emph{not} map to any region in the new reference genome (colored in pink). 
\item A \textit{new region} is a region in the new reference genome that does \emph{not} map to any region in the old reference genome (colored in green). 
\end{enumerate} 
We next describe how we identify and use these regions to quickly and
comprehensively remap a read set. 

\subsection{The AirLift Index}
\label{subsec:mech:airlift_index}

The \emph{AirLift Index} is comprised of two lookup tables (LUTs), each of
which has a one-time construction cost for any pair of reference genomes. The
LUTs describe regions of similarity between a pair of reference genomes, which
can then be used to quickly remap reads between the references. 

The first LUT, i.e., \emph{constant regions LUT}, associates each constant
region in the old reference with its respective region in the new reference
genome. AirLift queries this \emph{constant regions LUT} with a location
(of a previously-mapped read) in the old reference to quickly find a list of
corresponding locations in the new reference that have the same genome
sequence. AirLift uses this list of locations to update the mapping of the
read, as we explain in more detail in Section~\ref{subsec:airlifting_read}. 

The second LUT, i.e., \emph{updated regions LUT}, associates each updated
region in the old reference with its respective region in the new reference
genome.  AirLift queries this \emph{updated regions LUT} with a location
(of a previously-mapped read) in the old reference to quickly find a list of
corresponding locations in the new reference that have similar genome
sequences. AirLift uses this list of locations to update the mappings of the
read, as we explain in more detail in Section~\ref{subsec:airlifting_read}. 

Once constructed, the \emph{AirLift Index} is used to aid in the efficient
mapping of any number of reads from one reference genome to another reference
genome.  We store the LUTs as a BED file, which can then efficiently be queried to identify the label of each region given a range of positions and a chromosome. We next explain how to label regions in the reference and construct
the \emph{AirLift Index}. 

\subsection{Categorizing Regions of Similarity and Constructing the AirLift Index} 
\label{subsec:mech:data_structures} 

\begin{figure*}[htb]
    \centering
    \includegraphics[width=0.9\linewidth]{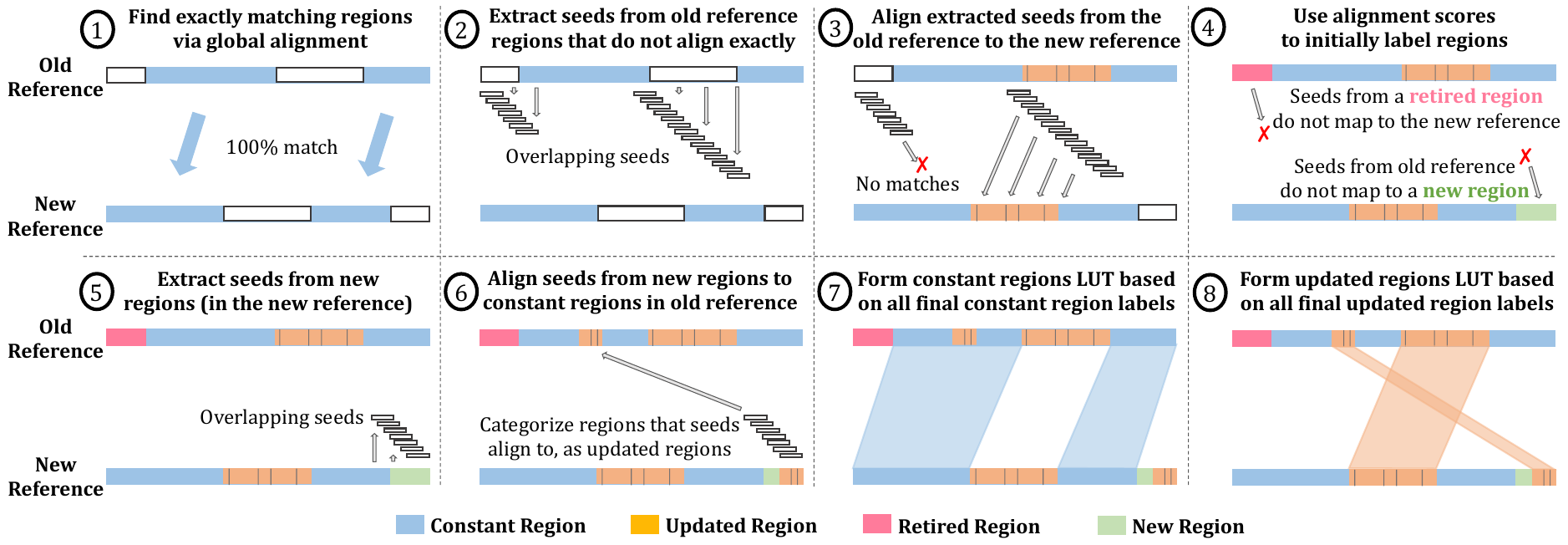}
    \caption{AirLift uses eight key steps to identify and label regions in the old and new reference genomes as \emph{constant}, \emph{updated}, \emph{retired}, or \emph{new} in order to efficiently map any number of reads from an old reference genome to a new reference genome.} 
    \label{fig:defining_regions}
\end{figure*}

The \emph{AirLift Index} is constructed via eight key steps, as we show in Figure~\ref{fig:defining_regions}. 

\textbf{(1)} First, we want to identify all regions (i.e., genome sequences)
that appear exactly the same in both the old and the new reference
genomes. To do so, we use a chain file (described in Supplementary
Section S2), which can be generated via BLAT~\cite{blat} with
exact matching (no errors allowed) global alignment. In
Figure~\ref{fig:defining_regions}, we indicate the constant regions in blue.

\textbf{(2)} In order to label the remaining regions in the new reference,
we first extract seeds (i.e., smaller subsequences) from regions in the old
reference that do \emph{not} map exactly to the new reference (non-blue
regions). Note that these seeds \textbf{a)} are the same length ($N$) as the
reads that we want to remap, and \textbf{b)} are overlapping seeds, i.e.,
completely overlap with each other such that a seed begins at each base pair
within each (non-blue) region and starting $N-1$ base pairs before each
(non-blue) region. This is to ensure that AirLift completely accounts for all
possible mapping locations, including sequences that may be partially included
in a constant region. 

\textbf{(3)} Next, we map the extracted seeds (from Step 2) to the new
reference genome to identify regions of approximate similarity across the
reference genomes. Note that this step can be done with \emph{any} read mapper.
We label as an \emph{updated region} (colored in orange) 1) any continuous
segment of base pairs that any seed has mapped to in the new reference or 2)
any continuous segment of seed locations in the old reference whose seeds have
mapped to the new reference. Since it is an approximate mapping, we indicate
differences between the updated regions in Figure~\ref{fig:defining_regions}
with black stripes. These differences are accounted for by the resulting
chain file. 

While we describe in more detail how we use these regions in
Section~\ref{subsec:airlifting_read}, we can quickly tell that if a read
mapped to an updated region in the old reference genome, there is a high chance
that the read will map to the respective updated region in the new reference
genome.  In order to comprehensively identify all possible locations in the new
reference that a read can map to just by examining the read's mapping
location in the old reference, we map seeds from the new reference using an
error rate of $2e$, where $e$ is the acceptable error rate for a successful
alignment, and report the best alignment. Due to our usage of a conservative error rate ($2e$), we are still
able to find every potential mapping with an alignment score within the
acceptable error rate (explained in Supplementary Section S4). 

\textbf{(4)} We find regions in the old reference where seeds (extracted from
Step 2) do \emph{not} align to and label them as \emph{retired regions}, since
the region or anything similar does not exist in the new reference genome.
Similarly, we find regions in the new reference whose seeds do not map to the
old reference genome and label them as \emph{new regions}, since the region or
anything similar to the region does not exist in the old reference genome.

\textbf{(5)} Next, we check to see whether regions within the new regions can be \emph{approximately} aligned to constant
regions in the old reference, since we had \emph{only} previously attempted mapping
seeds from non-constant regions to the new regions (in Step 3), and constant regions were
only identified with \textit{exact} matching. We do this by first extracting
overlapping seeds from the new regions.

\textbf{(6)} We then map the extracted overlapping seeds (from Step 5) to
the constant regions in the old reference genome. For any seeds that result in
a successful alignment, we 1) additionally label the corresponding segment
of the constant region as an updated region and 2) relabel the corresponding
segment of the new region as an updated region. We can now consider each of
these regions as updated regions, since this step has resulted in identifying
an associated similar region in the other reference. This step is necessary to
ensure that all regions in the old reference are checked for similarity to all
regions in the new reference, enabling a comprehensive mapping for reads that
map to \emph{any} region in the old reference. 

\textbf{(7)} We show the associated constant regions between the two
references within the areas shaded in blue and use this information to
create a \emph{constant regions LUT}, which can be queried with a location in
the old reference to obtain locations in the new reference that contain the
exact same sequence. We encode the mapping with the chain file format
(described in Supplementary Section S2). 

\textbf{(8)} We show the associated updated regions between the two
references within the areas shaded in orange and use this information to
create the \emph{updated regions LUT}, which can be queried to immediately
return candidate locations in the new reference that a read should be aligned
to. We encode the mapping and account for the minor differences using the chain
file format. 

\subsection{Using AirLift to Remap a Read} 
\label{subsec:airlifting_read}

\begin{figure}[ht]
    \centering
    \includegraphics[width=\linewidth]{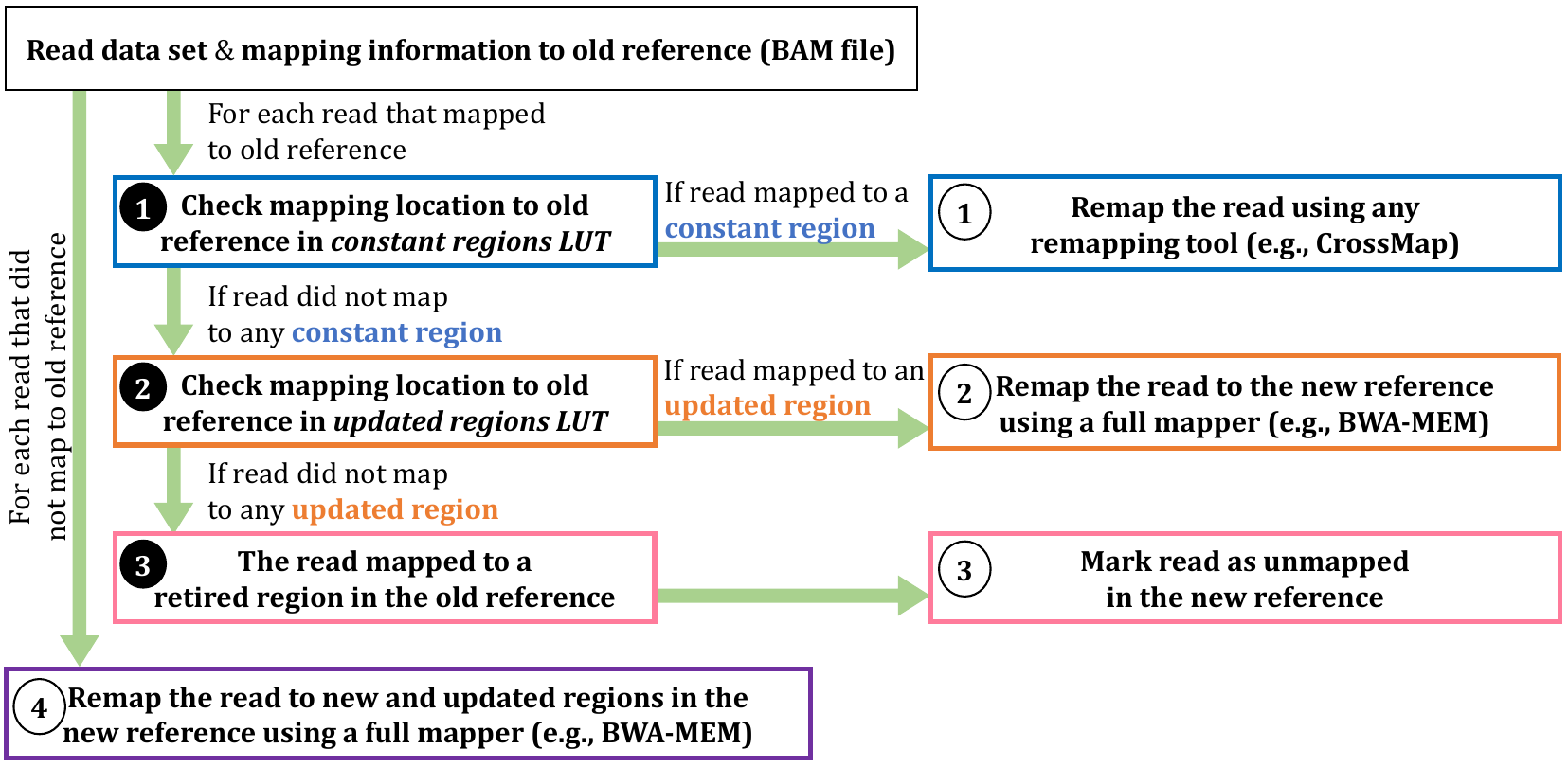}
    \caption{Using AirLift to remap a read set. AirLift remaps each read differently depending on the label of the region in the old reference that the read had originally mapped to: constant, updated, retired, or unmapped.}
    \label{fig:remap_cases}
\end{figure}

AirLift follows the procedure illustrated in Figure~\ref{fig:remap_cases}
to comprehensively and accurately remap a read set. AirLift first identifies
the label of the region that each read had originally mapped to in the old
reference using a series of steps (described in
Section~\ref{subsubsec:determining_label}). Depending on the label, AirLift
remaps each read using one of four independent cases (described in
Section~\ref{subsubsec:remapping_cases}), depending on the label of the region
that the read originally mapped to within the old reference: \textbf{(1)} a
read that mapped to a \emph{constant region}, \textbf{(2)} a read that mapped
to an \emph{updated region}, \textbf{(3)} a read that mapped to a \emph{retired
region}, and \textbf{(4)} a read that \emph{never mapped} to any location in
the old reference genome (i.e., an unmapped read). 

\subsubsection{Determining how to Remap each Read}
\label{subsubsec:determining_label}

To determine which case AirLift should apply when remapping a read, AirLift
performs the following steps on each read in the read set that originally
mapped to any location in the old reference. First, AirLift checks the read's
mapping location to the old reference in the \emph{constant regions LUT}
(\blackcircled{1} in Figure~\ref{fig:remap_cases}). If the mapping
location returns an associated location in the new reference, the read had been
originally mapped to a constant region in the old reference, and AirLift remaps
the read via Case \circled{1} (described in
Section~\ref{subsubsec:remapping_cases}).

If the \emph{constant regions LUT} does not return a location in the new
reference, AirLift next checks the read's mapping location to the old reference
in the \emph{updated regions LUT} (\blackcircled{2} in
Figure~\ref{fig:remap_cases}). If the mapping location returns an associated
location in the new reference, the read had been originally mapped to an
updated region in the old reference, and AirLift remaps the read via Case
\circled{2} (described in Section~\ref{subsubsec:remapping_cases}). 

If the \emph{updated regions LUT} does not return a location in the new
reference, the read had been originally mapped to a retired region in the old
reference (\blackcircled{3} in Figure~\ref{fig:remap_cases}).  This is
because an old reference is only comprised of constant, updated, and retired
regions, and AirLift already determined that the read was not originally mapped
to a constant or updated region. AirLift handles such reads via Case
\circled{3} (described in Section~\ref{subsubsec:remapping_cases}). 

In order to be comprehensive in remapping a read set, AirLift also
considers the reads that were unmapped in the old reference and attempts to
remap them to the new reference using Case \circled{4} (described in
Section~\ref{subsubsec:remapping_cases}). 

\subsubsection{Remapping each Read} 
\label{subsubsec:remapping_cases}

\textbf{Case 1:} For a read that had originally mapped to a \textit{constant
region}, we simply translate the mapping locations according to the offset in
the specific constant region from the old reference to the new reference. Since
this is the extent of existing state-of-the-art remapping tools' capabilities,
we can perform this step with any of these tools (e.g., \emph{LiftOver},
\emph{CrossMap}) for any read that is fully encapsulated within a chain file
interval. For our analysis, we built a new tool based on
CrossMap that is publicly released called FastRemap~\cite{kim2022fastremap, FastRemaptool}\footnote{FastRemap implements necessary modifications to the
\emph{CrossMap} code such that its output is compatible with GATK (See
Supplementary Section S5).}, that outputs BAM files
which can be used for downstream analysis (e.g., variant calling) for
validating our results. The chain file represents only regions that are exact
matches, so remapped reads will perfectly match to regions in the new reference
genome as well.

\textbf{Case 2:} For a read that maps to an \textit{updated region}, we first
query the \emph{updated regions LUT} to quickly obtain a list of locations
in the new reference genome that are similar (within a $2e$ error rate) to the
location that the read mapped to in the old reference genome.  We can then use
any aligner (e.g., BWA-MEM~\cite{li2013aligning}) to align the read to all locations returned by the
\emph{updated regions LUT} and return the locations in the new reference
genome that align with an error rate smaller than a user-defined error rate
($e$). 

\textbf{Case 3:} For a read that maps to a \textit{retired region} (in the old
reference genome), we already know that the read will not map anywhere in the
new reference genome, since retired regions are not similar to any region in
the new reference genome. Therefore, we can mark that read as an unmapped read
in the new reference genome.

\textbf{Case 4:} For a read that \textit{never mapped anywhere} in the old
reference genome, we know that the read will not map to any constant region in
the new reference genome. However, there is a chance that the read can align to
updated or new regions in the new reference genome. Therefore, we must fully
map the read to each new and updated region using any read mapper.

\section{Evaluation} \label{sec:eval}
\subsection{Evaluation Methodology} \label{subsec:evmethod}

\textbf{AirLift Tools.} AirLift uses 1)~FastRemap~\cite{FastRemaptool,kim2022fastremap}, a recent tool
based on \emph{CrossMap}~\cite{zhao2013crossmap, crossmap}, to quickly move all reads that map to constant regions in the old reference, 2)~\emph{BWA-MEM}~\cite{li2013aligning} to map reads when constructing the
\emph{AirLift Index} to identify the regions in both old and new reference genomes and 3)~\emph{BWA-MEM}~\cite{li2013aligning} for mapping the reads that are not in the constant regions identified with the \emph{AirLift Index}.
 
\noindent\textbf{Evaluated Remappers.} We evaluate two widely used
remappers, \emph{CrossMap}~\cite{zhao2013crossmap,crossmap} and \emph{UCSC
LiftOver}~\cite{ucscliftover} to compare against AirLift. Note that these two
remappers do \emph{not} provide a \emph{comprehensive} or \emph{accurate}
solutions to remapping reads from one reference to another. Due to the
limitations of prior remappers (described in Supplementary
Section S3), we evaluate and compare against the only comprehensive and accurate baseline of \emph{fully mapping} the read set (from scratch without using any prior mapping information) to the new reference genome with \emph{BWA-MEM}~\cite{li2013aligning}.

\noindent\textbf{Evaluated Reference Genomes Read Data Sets.} We evaluate AirLift with several versions of reference genomes of varying size across 3 species (i.e., human, C. elegans, yeast) as shown in Supplementary Table S2. 
We use DNA-seq read sets from four
different samples of the set of species whose reference genomes we examine (as
shown in Supplementary Table S3). 

\noindent\textbf{Evaluating Accuracy.} To evaluate the accuracy, we perform variant calling by using the mapping information from 1)~AirLift and 2)~fully mapping from scratch. For variant calling, we use \emph{GATK
HaplotypeCaller}~\cite{McKenna2010} by following the best
practices~\cite{Auwera2013} and \jkt{VCFtools~\cite{danecek2011variant} to filter variant calling files based on a minimum quality score of 30 (i.e., $--$minQ
30)}.
To benchmark the variant calling results, we use the \emph{hap.py} tool (https://github.com/Illumina/hap.py). 

\noindent\textbf{Evaluation System.} We run AirLift on a server with 64 cores
(2 threads per core, AMD EPYC 7742 @ 2.25GHz), and 1TB of the memory. We assign
32 threads for C. elegans and yeast and 48 threads for human genomes when
running tools with multithreaded capabilities (i.e., SAMtools, BWA-MEM;
described in Supplementary Section S7) and collect their
runtimes (usr and sys) and memory usage using the \texttt{time} command in
Linux with \texttt{-vp} flags. We report the aggregate runtime (in
seconds) and peak memory usage (in megabytes) across all active threads in our
evaluations with these configurations.

\begin{figure*}[!t]
    \centering
    \includegraphics[width=\linewidth]{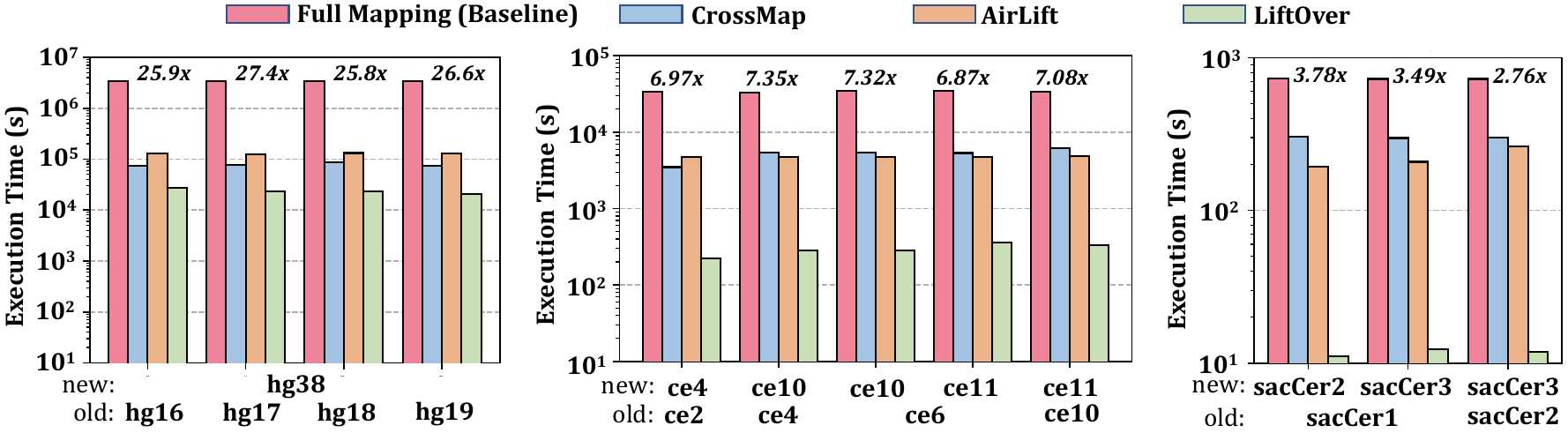}
    \caption{\textbf{AirLift execution time results.} We show the execution time (log-scale y-axis) of running three remapping tools, \emph{CrossMap} (blue), AirLift (orange), and \emph{LiftOver} (green) on a read set to a new reference genome against the baseline (red) of \emph{fully mapping} a read set to the new reference genome. We plot the execution times of each tool for various pairs of reference genomes (x-axis; where the old reference is at the bottom and the new reference is above the old reference) in three separate plots for different sizes of reference genomes, i.e., large (human), medium (C. elegans), small (yeast). We indicate the speedup of AirLift against the \emph{full mapping} baseline above each grouping of bars, since AirLift and the baseline are the only comprehensive and accurate remapping techniques available.}
    \label{fig:airlift_speedup}
\end{figure*}

\noindent\textbf{AirLift Evaluation Plots.} 
In each AirLift evaluation plot, we show on the x-axis, both the old reference
genome (below) and the new reference genome (above) used in the evaluation.
Note that in our evaluations of AirLift, we \emph{only} consider the remapping
stage (as other stages are \emph{preprocessing stages} that are performed once for
each pair of reference genomes for building the AirLift Index). We show the execution times and memory usage of the
preprocessing stage in Supplementary Tables S4 and S5.

\subsection{AirLift Execution Time} 
\label{subsec:remap_results}

We first demonstrate how AirLift reduces the time to map a set of reads to an
updated reference genome by reducing the number of reads that we must map.
Figure~\ref{fig:airlift_speedup} plots the execution times (y-axis) for mapping
a read set to a new reference genome using three different remapping tools,
\emph{CrossMap}, AirLift, and \emph{LiftOver} compared to the baseline of
\emph{fully remapping} the entire read set from an old reference genome to the
new reference genome.  We provide the speedup of AirLift over fully mapping the
read set to the new reference (i.e., $T_{\text{Full
Mapping}}/T_{\text{AirLift}}$) above each bar. 

The execution time of AirLift is calculated as the sum of the execution times
for performing each of the cases (described in
Section~\ref{subsubsec:remapping_cases}) as shown in Equation~\ref{eq:time},
where $T_{\text{constant}}$ is the time to translate all reads that
originally map to a constant region in the old reference,
$T_{\text{updated}}$ is the time to map all reads that originally mapped
to an updated region in the old reference, $T_{\text{retired}}$ is the
time to map all reads that are originally mapped to a retired region in the old
reference, and $T_{\text{unmapped}}$ is the time to map all reads that never
mapped anywhere in the old reference. The exact execution time breakdowns for
each of these four cases are shown in Supplementary
Table S7. We also provide the number and ratio of reads that
AirLift must remap in each case for each pair of references in Supplementary
Tables S8 and S9, respectively, and the average time per read per case for each
pair of references in Supplementary Table S10. 
\begin{equation}\label{eq:time}
T_{\text{AirLift}} = T_{\text{constant}} + T_{\text{updated}} + T_{\text{retired}} + T_{\text{unmapped}}
\end{equation}

We make three observations based on Figure~\ref{fig:airlift_speedup} and the
supplementary tables. 
First, AirLift consistently provides significant speedup over the baseline
(of \emph{fully mapping} a read set) across all tested pairs of references,
ranging from $2.76\times$ (\pair{sacCer2}{sacCer3}) up to $27.4\times$
(\pair{hg17}{hg38}). This is because the AirLift execution time is largely comprised of the
time to remap the reads within the constants regions (i.e., between 86.57\% for \pair{hg16}{hg38} and 98.47\% for \pair{ce10}{ce11}), which is more efficient than fully remapping a read from scratch to the entire reference genome.
Second, remapping a read set
with AirLift between a pair of references with a smaller constant region size
results in a higher execution time. Therefore, AirLift performs faster when
remapping reads between pairs of references that are more similar to each
other.
Third, AirLift is slightly faster than CrossMap for some reference genome pairs (e.g., \pair{ce4}{ce10}), although AirLift includes additional steps than CrossMap, such as fully remapping reads from scratch to the entire reference genome. This is still expected as AirLift uses FastRemap when remapping reads within constant regions, which is substantially faster than CrossMap~\cite{kim2022fastremap}. We conclude that AirLift substantially improves the performance of remapping reads comprehensively compared to full mapping.

We conclude that AirLift significantly improves the execution time for
\emph{comprehensively} and \emph{accurately} remapping a read set from an old
reference to a new reference compared to the baseline of \emph{fully mapping}
the read set to the new reference.

\subsection{AirLift Memory Usage} 
\label{subsec:memory_results}

\begin{figure*}[!b]
    \centering
    \includegraphics[width=\linewidth]{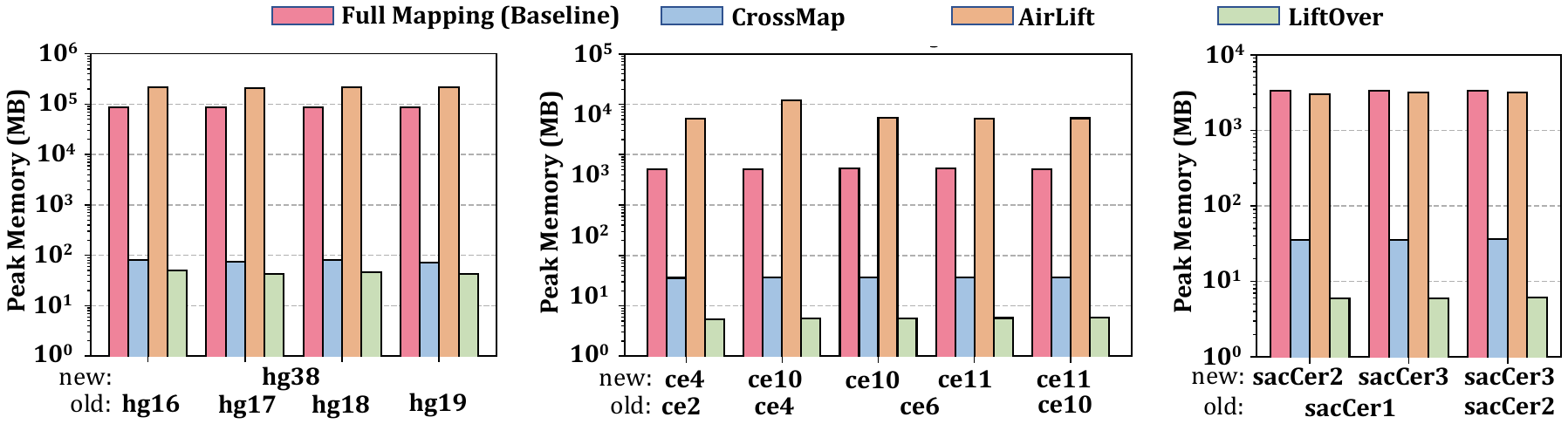}
    \caption{\textbf{AirLift memory usage results.} Peak memory usage results for each of the remapping tools during remapping.} 
    \label{fig:airlift_memory} 
\end{figure*}

Figure~\ref{fig:airlift_memory} plots the peak memory usage in MB (y-axis)
across the remapping tools (i.e., \emph{CrossMap}, AirLift, and
\emph{LiftOver}) and baseline full mapping method (i.e., \emph{BWA-MEM}) for
our set of evaluated reference pairs (x-axis). We find that across all tested
reference pairs, AirLift has similar peak memory requirements as our \emph{full
mapping} baseline, \emph{BWA-MEM}.  This is because AirLift relies on
\emph{BWA-MEM} to remap a portion (i.e., up to 16.61\%) of the read set, which
is large enough to require the same amount of memory as mapping the full read
set. 

\subsection{GATK Variant Calling Results} 
\label{subsec:gatk_results} 

To demonstrate that AirLift provides similar mapping results as a \emph{full
mapper} (baseline) and it is much more comprehensive and accurate than
\emph{CrossMap} and \emph{LiftOver}\footnote{The GATK HaplotypeCaller tool
cannot analyze the outputs of \emph{CrossMap} or \emph{LiftOver} since their
outputs are not compatible with downstream analysis tools (as described in
Supplementary Section S5). Therefore, we do not analyze
the outputs of \emph{CrossMap} or \emph{LiftOver} in this section.}, we
perform downstream analysis (i.e., variant calling). We use the GATK
HaplotypeCaller tool to call variants from both the 1) \emph{full mapping} BAM
file and 2) Airlift-generated BAM file. We use the VQSR~\cite{VQSR} tool 
to recalibrate the variants based on quality scores provided by the GATK
HaplotypeCaller tool. We use the hap.py tool to benchmark 1) the AirLift
variant calls against full mapping, 2) the AirLift variant calls against the
gold standard (i.e., ground truth), and 3) full mapping variant calls against
the ground truth, if the ground truth is available. We use the variant calling
ground truth from the Platinum Genomes~\cite{Eberle2017} and Genome in a
Bottle~\cite{Zook2014} for the human NA12878 sample. We only benchmark Airlift
against \emph{full mapping} for the C. elegans and yeast data sets, since we do
not have the ground truth for these species. We report the precision and recall
results for the SNPs and insertion-deletions (indels) as calculated by hap.py
(https://github.com/Illumina/hap.py). 

Table~\ref{tab:vcalln} shows the variant calling results for human, C. elegans,
and yeast genomes, respectively.  Each row contains quality measurements of
identifying single nucleotide polymorphisms (SNPs) and insertion-deletions
(indels) for a pair of reference genomes in terms of precision and recall
(written as `precision score(\%)/recall score(\%)'). For the human results, we
show the precision and recall scores of \emph{full mapping} when
identifying the set of SNPs and indels compared against the set of SNPs and
indels that the ground truth reports, to demonstrate how AirLift compares
against \emph{full mapping} when identifying ground truth SNPs and indels.  The
columns are separated to show separate precision and recall scores for
identifying the set of SNPs and indels when compared against the set of SNPs
and indels that \emph{full mapping} identifies (vs. Full Mapping) and the
ground truth reports (vs. Ground Truth; only available for human results).

\begin{table*}[ht]
    \footnotesize 
    \caption{GATK Variant Calling Results for Human, C. elegans, and Yeast Genomes} 
    \begin{center}
    \begin{tabular}{rll||cccc} 
	\textbf{Remap} & \multicolumn{2}{c}{\textbf{Read Sets}} & \multicolumn{2}{c}{\textbf{vs. Full Mapping}} & \multicolumn{2}{c}{\textbf{vs. Ground Truth}} \\ 
	\textbf{Technique} & from & to & SNP (\%) & Indel (\%) & SNP (\%) & Indel (\%) \\ 
	\hline 
    \hline
 	
	\multirow{1}{*}{Full Mapping} & \multicolumn{1}{c}{-} & hg38             & - & - & 97.73/99.25 & 81.46/96.27 \\
	\hline
	\multirow{4}{*}{AirLift} & hg16 & \multirow{4}{*}{hg38} & 95.00/97.87 & 75.20/91.03 & 96.65/95.79 & 84.26/88.13 \\ 
	                         & hg17 &                       & 95.16/98.12 & 75.59/91.30 & 96.97/95.82 & 84.72/88.42 \\ 
	                         & hg18 &                       & 95.50/98.22 & 75.76/91.38 & 96.93/96.05 & 85.01/88.75 \\ 
	                         & hg19 &                       & 95.99/98.38 & 76.12/91.51 & 97.05/96.39 & 85.50/89.02 \\ 
	\hline
	
	\multirow{5}{*}{AirLift} & ce2  & ce4                   & 90.82/97.29 & 96.97/97.66 & - & - \\ \cline{2-7}
	                         & ce4  & \multirow{2}{*}{ce10} & 91.06/96.96 & 96.81/97.30 & - & - \\
	                         & ce6  &                       & 91.11/97.00 & 96.81/97.33 & - & - \\ \cline{2-7}
	                         & ce6  & \multirow{2}{*}{ce11} & 90.01/96.12 & 95.86/96.18 & - & - \\ 
	                         & ce10 &                       & 90.03/96.48 & 95.90/96.44 & - & - \\ 
	\hline 
	\multirow{3}{*}{AirLift} & sacCer1 & sacCer2                  & 95.30/98.82 & 95.83/94.74 & - & - \\ \cline{2-7}
	                         & sacCer1 & \multirow{2}{*}{sacCer3} & 86.35/94.27 & 90.38/88.65 & - & - \\ 
	                         & sacCer2 &                          & 87.03/91.19 & 91.14/88.65 & - & - \\ 
	\hline 
	\end{tabular} 
    \end{center}
    \label{tab:vcalln}

    \begin{minipage}{\textwidth}
    \footnotesize
GATK results of the read sets from all evaluated species remapped by AirLift from an older reference version (e.g., hg16, hg17) to a more recent reference version (e.g., hg38) and for fully mapping (via BWA-MEM) the read set to the latest human reference version (since we only have ground truth GATK values for the human reference). For each read set remapped by AirLift, we show the precision(\%)/recall(\%) results of identifying SNPs and indels compared to 1) \emph{full mapping} and 2) the ground truth. We also show the results of fully mapping the read set to hg38 compared to the ground truth. All results were obtained using GATK HaplotypeCaller~\cite{McKenna2010} and hap.py. 
    \end{minipage}
     
\end{table*}

We make two key observations. First, we observe that AirLift is able to
identify SNPs reported by \emph{full mapping} with high precision and recall
scores (as shown under the first column, \emph{vs. Full Mapping}).  This is
because AirLift 1) identifies all possible mapping locations for each read in
the read set similarly to the \emph{full mapping} approach, 2) comprehensively
maps each read accordingly, and 3) reports accurate alignment results (i.e.,
alignments with error rates below a specified acceptable error rate) unlike
existing remapping tools. Second, we observe that AirLift identifies SNPs and
indels reported by the ground truth with precision similar to \emph{full
mapping}. We observe this by comparing the results in the first row (i.e.,
\emph{Full Mapping}) against the AirLift results directly underneath them
(e.g., 97.73\%/99.25\% precision/recall values for identifying SNPs when
full mapping to hg38 compared to 97.05\%/96.39\% when using AirLift
between \pair{hg19}{hg38}; only available for human results).

We note the small variation across precision and recall values in the table and attribute them to
two main factors. The first factor is due to the mapping quality scores that are recalculated when fully mapping them. Fully recalculating mapping scores can enable improving the overall accuracy of the mapping quality in an alignment, which improves the overall accuracy in variant calling~\cite{langmead_tandem_2017}. However, AirLift does not recalculate the mapping quality scores of the reads that fall under the constant regions for fast mapping, which may result in non-optimal mapping and mapping quality score calculation in case these reads map to updated or new regions with a better mapping quality score. This is a trade-off between using an efficient and highly accurate remapping tool, AirLift, and fully mapping reads with slightly higher accuracy, which comes with significantly higher computational costs.
The second factor is due to the discrepancies that may
occur as a result of genomic repeats and reproducibility issues in
BWA-MEM~\cite{firtina2016}. We argue that these alignment differences do not
cause a significant loss in variant calling quality, as AirLift precision and
recall results for SNPs and indels are very similar to full mapping (when both
are benchmarked against the ground truth).

We have shown in our evaluations against existing state-of-the-art remapping
tools, that AirLift can comprehensively and accurately remap a read set from
one reference genome to another at high speeds (i.e., up to $27.4\times$
faster than our \emph{full mapping} baseline). Since AirLift accomplishes our
four goals of remapping a read set quickly, comprehensively, accurately, and
end-to-end, providing a BAM-to-BAM result that can be immediately used in
downstream analysis, we conclude that AirLift is a viable tool to be used as a
quick alternative to fully mapping a read set when it had previously been
mapped to a similar reference genome.

\section{Conclusion} 

We introduce AirLift, a methodology and tool for quickly, comprehensively,
and accurately remapping a read data set that had previously been mapped to an
older reference genome to a newer reference genome. AirLift is the first tool
that provides BAM-to-BAM remapping results of a read data set on which
downstream analysis can be immediately performed. The key idea of AirLift is to
construct and use an \emph{AirLift Index}, which exploits the similarity between
two references to quickly identify candidate locations that a read should be
remapped to based on its original mapping in the old reference. We compare
AirLift against several existing remapping tools, CrossMap and LiftOver, which
we demonstrate have several major limitations. These tools either do
\emph{not} provide accurate and comprehensive remapping results or do not
result in remapping results on which downstream analysis can be immediately
performed (summarized in Table~\ref{tab:prior_dram_works}). We compare
AirLift against the only comprehensive and accurate method of \emph{fully
mapping} a read data set to the new reference using BWA-MEM, and find that
AirLift significantly reduces the execution time by up to 27.4$\times$, 7.35$\times$, and 3.78$\times$ for large (human), medium (C. elegans), and
small (yeast) reference genomes, respectively. We validate our results against
the ground truth and show that AirLift identifies similar rates of SNPs and
Indels as the full mapping baseline. We conclude that AirLift is the first
comprehensive and accurate remapping tool that substantially reduces the
execution time of remapping a read data set, while providing end-to-end
BAM-to-BAM results on which downstream analysis can be performed. We look
forward to future works that take advantage of as well as improve AirLift for
various genomic analysis studies.

\section{Conclusion} 

We introduce AirLift, a methodology and tool for quickly, comprehensively,
and accurately remapping a read data set that had previously been mapped to an
older reference genome to a newer reference genome. AirLift is the first tool
that provides BAM-to-BAM remapping results of a read data set on which
downstream analysis can be immediately performed. The key idea of AirLift is to
construct and use an \emph{AirLift Index}, which exploits the similarity between
two references to quickly identify candidate locations that a read should be
remapped to based on its original mapping in the old reference. We compare
AirLift against several existing remapping tools, CrossMap and LiftOver, which
we demonstrate have several major limitations. These tools either do
\emph{not} provide accurate and comprehensive remapping results or do not
result in remapping results on which downstream analysis can be immediately
performed (summarized in Table~\ref{tab:prior_dram_works}). We compare
AirLift against the only comprehensive and accurate method of \emph{fully
mapping} a read data set to the new reference using BWA-MEM, and find that
AirLift significantly reduces the execution time by up to 27.4$\times$, 7.35$\times$, and 3.78$\times$ for large (human), medium (C. elegans), and
small (yeast) reference genomes, respectively. We validate our results against
the ground truth and show that AirLift identifies similar rates of SNPs and
Indels as the full mapping baseline. We conclude that AirLift is the first
comprehensive and accurate remapping tool that substantially reduces the
execution time of remapping a read data set, while providing end-to-end
BAM-to-BAM results on which downstream analysis can be performed. We look
forward to future works that take advantage of as well as improve AirLift for
various genomic analysis studies.

\section*{Data Availability}

The Human NA12878 illumina read data set is publicly available (Accession
number ERR194147 and ERR262997). The C. elegans N2 illumina read data set is
publicly available (Accession number SRR3536210). The Yeast S288C illumina read
data set is publicly available (Accession number ERR 1938683).

\balance

\setstretch{0.8}
\bibliographystyle{IEEEtran}
{\small \bibliography{main}}
\setstretch{1}
\clearpage
\onecolumn
\renewcommand\ltitle{AirLift: A Fast and Comprehensive Technique \\ for Remapping Alignments between Reference Genomes\xspace}
\begin{center}
    \LARGE{Supplementary Text for\\\ltitle}
\end{center}

\setcounter{section}{0}
\setcounter{equation}{0}
\setcounter{figure}{0}
\setcounter{table}{0}
\setcounter{page}{1}

\renewcommand{\theequation}{S\arabic{equation}}
\renewcommand{\thetable}{S\arabic{table}}
\renewcommand{\thefigure}{S\arabic{figure}}
\renewcommand{\thesection}{S\arabic{section}}
\renewcommand{\thetheorem}{S\arabic{theorem}}

\section{Currently Available Remapping Tools}
\label{supp:remapping_tools}

\noindent\textbf{UCSC LiftOver.} One of the most commonly used remapping tools
is UCSC LiftOver~\protect\citesupp{supp_ucscliftover}. UCSC LiftOver uses a chain file between
two different assemblies of a genome to convert the coordinates from one
assembly to the assembly of the other genome. UCSC LiftOver suffers from three
major shortcomings. First, UCSC LiftOver functionality is limited to the
genomes whose assemblies are provided by the UCSC Genome
Browser~\protect\citesupp{supp_ucscgenomebrowser}, hence, making it impossible to remap genomes
whose assemblies are not yet included in the tool. Second, the tool \emph{only}
converts the coordinates of regions within the old reference genome that are
highly similar to regions within the updated reference genome and ignores
regions with significant variance, which prevents a comprehensive remapping of the
coordinates (described in more detail in Supplementary
Section~\ref{supp:limitations}). Third, UCSC LiftOver only supports \rev{the} BED-format
(i.e., browser extensible data) input files which limits its usage even
further.

\noindent\textbf{CrossMap.} One alternative to UCSC LiftOver is
CrossMap~\protect\citesupp{supp_zhao2013crossmap, supp_crossmap}. CrossMap follows a similar approach
with UCSC LiftOver and uses chain files to convert mappings from an older
reference genome to a newer reference genome. Compared to UCSC LiftOver,
CrossMap supports a larger set of input file formats, such as BAM, SAM, or
CRAM, BED, Wiggle, BigWig, GFF (i.e., general feature format) or GTF (i.e.,
gene transfer format), and VCF (i.e., variant call
format)~\protect\citesupp{supp_zhao2013crossmap, supp_crossmap}. Unfortunately, CrossMap suffers from
similar limitations as UCSC LiftOver. 

\noindent\textbf{NCBI Genome Remapping Service.} Another alternative is NCBI
Genome Remapping Service~\protect\citesupp{supp_ncbi_genome_remap}, which also remaps the
annotations from one genome assembly to another. NCBI Remap has support for a
larger set of input/output file formats, such as BED, GFF, GTF, and VCF. NCBI
Remap can also perform cross species remapping for a limited number of
organisms. However, as with UCSC LiftOver, NCBI Remap is limited \rev{to} the
provided assemblies. 

\noindent\textbf{Segment\_liftover.}
Segment\_liftover~\protect\citesupp{supp_gao2018segment_liftover, supp_segmentliftover} is another
tool that is designed to map coordinates of one genome assembly to another
genome's assembly while maintaining the integrity of the genome segments that
are not continuous anymore in the target assembly. However, Segment\_liftover
first runs UCSC LiftOver and then attempts to approximately map any failed
conversions~\protect\citesupp{supp_gao2018segment_liftover}. Due to the high coverage of UCSC
LiftOver in remapping segments, most conversions are performed by UCSC LiftOver
and therefore suffer from the same shortcomings of UCSC LiftOver. 

\noindent\textbf{Galaxy.} Galaxy~\protect\citesupp{supp_giardine2005galaxy, supp_galaxy_remap} is a
web-based platform, which has LiftOver as part of its toolset. This tool is
based on UCSC LiftOver~\protect\citesupp{supp_ucscliftover} and the chain files provided by UCSC
Genome Browser~\protect\citesupp{supp_ucscgenomebrowser}. Thus, Galaxy also suffers from similar
limitations as UCSC LiftOver. 

\noindent\textbf{PyLiftover.} PyLiftover~\protect\citesupp{supp_pyliftover_remap} is a Python
implementation of a limited version of UCSC LiftOver. PyLiftover does not
convert ranges (i.e., only converts point coordinates) between different
assemblies, and it does not support BED-format input files.

\noindent\textbf{Bazam.} Bazam~\protect\citesupp{supp_sadedin2019bazam} is another tool which remaps short paired reads by optimizing memory usage while providing high parallelism. However, Bazam \emph{only} targets the steps where reads are read from a BAM or CRAM file (i.e., read extraction) and sent to an aligner (e.g., BWA~\protect\citesupp{supp_li2009fast}). Eventually, \emph{all} the reads are remapped to the new reference genome, which is inefficient. %

\noindent\textbf{nf-LO.} nf-LO~\protect\citesupp{supp_talenti2021nf} is a containerized implementation of UCSC LiftOver written in Nextflow, which enables its usage in any Unix-based system. However, as nf-LO is directly based on UCSC LiftOver, it comes with the same limitations.

\noindent\textbf{LevioSAM.} LevioSAM~\protect\citesupp{supp_mun_leviosam_2021, supp_chen_improved_2022} is a remapping tool that remaps reads from a variant-aware reference to another reference using a VCF file or a chain file. LevioSAM creates a separate index file for querying remapping and updating the CIGAR string of remapped reads. AirLift provides an efficient and comprehensive lookup for reads that can be remapped without updating the CIGAR string and sends the remaining reads to a read mapper to perform a full read mapping, which can align these reads to better regions with updated CIGAR strings.

\noindent\textbf{Liftoff.} Liftoff~\protect\citesupp{supp_shumate2021liftoff} uses similar methods as remapping tools, but focuses primarily on remapping genes between two references. AirLift on the other hand remaps the full read set between two references, providing coverage on non-coding regions of the genome that may contain important information (e.g., gene expression).

\section{Chain File Format} 
\label{supp:chain_file}

The chain file~\protect\citesupp{supp_chainformat} is a commonly used data structure across
remapping tools and essentially describes the relationship of two reference
genomes. The chain file is typically generated with two steps: 1) performing
global alignment to detect similar regions between two reference genomes, and
2) encoding the identified similarities into a simple readable format.  The
chain file encodes the differences of large genomic sequences (e.g.,
chromosomes) as a list of three-integer-tuples. The first integer represents
the length of the \emph{alignment strand}, or shared sequence. The second
integer represents the length of the \emph{gap}, or different sequence in the
old reference genome.  The third integer represents the length of the gap in
the new reference genome.  In this way, the offset of an alignment strand
across the old and new reference genomes can be quickly calculated, and reads
that fall within the alignment strand can be quickly remapped according to the
offset. 

\section{Limitations of Currently Available Remapping Tools} 
\label{supp:limitations}

Repeating a genomic study using a different version of the reference genome is
computationally very expensive.  A faster and more convenient way to achieve
this is to ``remap'' the mapping locations from the older reference genome to
its updated version~\protect\citesupp{supp_ucscliftover, supp_crossmap, supp_segmentliftover,
supp_gao2018segment_liftover, supp_zhao2013crossmap, supp_ncbi_genome_remap, supp_galaxy_remap,
supp_pyliftover_remap}. While these tools (described in
Section~\ref{supp:remapping_tools}) can quickly move many annotations, there
are several limitations with the current methodology that we study and
demonstrate with UCSC LiftOver~\protect\citesupp{supp_ucscliftover}. UCSC LiftOver is both the
state-of-the-art tool commonly used for remapping reads and also the codebase
wrapped or modeled by several other tools~\protect\citesupp{supp_segmentliftover,
supp_giardine2005galaxy, supp_galaxy_remap, supp_pyliftover_remap}. In Figure~\figlimitations,
read 2 (mapped to location 1210 in the old reference) shows an example of how a
read is remapped using such a tool. The tool identifies a region corresponding
to the region that read 2 maps to (region A) that is similar across the two
reference. Since the region begins at location 1180 in the new reference and to
1175 in the old reference, all reads mapping to region B are also shifted by 5
base pairs when remapping them to the new reference. 

In our evaluation of UCSC LiftOver~\protect\citesupp{supp_ucscliftover}, we find that
techniques relying on the standard chain file format do not account for
large insertions (i.e., many new base pairs that exist in the new reference but
not in the old reference) in the new reference genome. Discounting insertions
results in two problems when using these techniques: 1) a remapped read can
contain a large insertion in the new reference resulting in a poor alignment
and low accuracy, and 2) insertions have low coverage in the new reference due
to the limitations of chain files resulting in low coverage of the new
reference. We illustrate these issues in Figure~\figlimitations~
(Reads 3 and 1, respectively). 

\subsection{Limitation 1: Low Accuracy} 
\label{mot:subsec:limit1}

The first limitation we identify is that UCSC LiftOver~\protect\citesupp{supp_ucscliftover} only
accounts for the overlap between a read and alignment sequences in the old
reference genome when remapping the read. A read will be remapped to the new
reference genome if the total length of gaps in the old reference genome between
the start and end of the read is less than the read length multiplied by
the selected error acceptance rate (5\% is typically used in read alignment.
This corresponds to the \emph{Minimum ratio of bases that must remap} parameter
on the UCSC LiftOver webtool~\protect\citesupp{supp_ucscliftover} being set to 0.95). However,
the tool remaps the read \emph{regardless} of the total length of the gaps in
the new reference genome.  This means, that if there is a large insertion in
the new reference genome between the start and end of the read in the alignment
strand, the read will \emph{still} be mapped even if the read no longer aligns
to that location with an error acceptance rate of 5\%. For example, read 3 in
Figure~\figlimitations~maps to the old reference genome at
location 1360. While there are 4 base pairs of difference (1375-1379) at the
read's mapping location in the old reference, it is within the 5\% error
acceptance rate and therefore will be remapped to location 1365 in the new
reference. The new mapping in the new reference has an insertion of 20 base
pairs long, which means that the new mapping can have an error rate of 20\%
(which is well beyond the 5\% error acceptance rate). In our evaluation of UCSC
LiftOver (with an error acceptance rate of 5\%, and reads of length 100 base
pairs), 0.41\% of remapped reads resulted in an error rate greater than 5\%
(often times much greater, i.e., $>$40\%) when aligned to the sequence at the
remapped location in the new reference genome. 

\subsection{Limitation 2: Low Coverage} 
\label{mot:subsec:limit2}

The second limitation we identify is that UCSC LiftOver~\protect\citesupp{supp_ucscliftover} is
inherently unable to remap reads 1) to regions in the new reference genome with
large insertions (i.e., regions that do not appear in the old reference) or 2)
that map to deleted regions in the old reference (i.e., regions that do not
appear in the new reference). This results in low coverage of those regions in
the new reference. For example, Read 1 in Figure~\figlimitations~ maps to the
old reference genome in a deleted region (1030-1130). However, since the chain
file cannot relay how that region relates to the new reference, the read cannot
be moved to the new reference. In addition, the large insertion in the new
reference (1000-1180) does not get mapped to since reads never mapped to
regions similar to it in the old reference genome. To demonstrate the
implications of this limitation, we examine chain files to identify the
\emph{theoretical} minimum number of annotations that are missed due to the
limitation (i.e., any annotation that falls within regions in the new reference
that are not covered by the chain file). In Supplementary
Figure~\ref{fig:crossmap_analysis1}, we show the minimum amount of information lost
when remapping from one human reference genome version (x-axis) to the latest
human reference genome version (hg38).  The y-axis shows the minimum percentage
of annotations (labeled and marked with unique colors) missed when remapping
solely with existing chain files. We make two key observations based on
Supplementary Figure~\ref{fig:crossmap_analysis1}.  First, we observe that a
significant portion ($>$7\%) of the genes and transcripts are lost when simply
using an available remapping tool (i.e., UCSC LiftOver) between hg16 and hg38.
Second, the percentage of the missed annotations decreases as the difference in
versions becomes smaller, but even when lifting annotations between hg19
(released in 2009) and hg38 (released in 2013), 4.47\% of genes are ``lost'' in
hg38.  Supplementary Table~\ref{fig:crossmap_analysis1}~contains the exact values
of each lost annotation (in percentages) when using UCSC LiftOver from hg16,
hg17, hg18, and hg19 (rows) to hg19 and hg38 (columns). We expect to observe
similar behavior in tools that wrap UCSC LiftOver~(e.g., \protect\citesupp{supp_segmentliftover,
supp_giardine2005galaxy, supp_galaxy_remap, supp_pyliftover_remap}).

\subsection{The Need for a Comprehensive Remapping Tool}

As the output of lifting annotations from one reference to another is used in
downstream genome analysis, we argue that the speed and accuracy of lifting
annotations, and coverage of the new reference genome are \emph{all} crucial.
However, prior works mainly focus on the speed at the cost of both accuracy and
coverage.  These remapping tools are often very inaccurate and can only lift
mappings or annotations for regions with minor
changes~\protect\citesupp{supp_zheng2017alignment}.  Therefore, if researchers want a
comprehensive study using a new reference genome, they must map the entire read
data set to the new reference genome rather than rely on the results of such
remapping tools~\protect\citesupp{supp_zheng2017alignment}.  Due to the high similarity between
the old and new reference genomes, we can use information from the old mapping
to \emph{very quickly} map a read data set to an updated reference genome.
\textbf{Our goal} is to produce a method for quickly remapping the reads of a
sample from one reference genome to an updated version of the reference genome
or another similar reference genome with high genome coverage. 

\section{Creating the AirLift Index with a $2e$ Error Rate}\label{suppsec:errorrate}

In Supplementary Figure~\ref{fig:remap_2e}, we illustrate an
example of why using a $2e$ error rate enables AirLift to find all possible
alignments of a read in the old reference. In Supplementary
Figure~\ref{fig:remap_2e}, a read (of length 20) aligns to a subsequence in the
updated region of the old reference genome with an $e=5\%$ error rate (one
mismatch on the $9^{th}$ base pair), and also aligns to a subsequence in the
updated region of the new reference genome with an $e=5\%$ error rate (one
mismatch on the $16^{th}$ base pair). While the read could map to either of the
regions with a 5\% (e) error rate, the sequences between the updated regions
exhibit a 10\% (2e) error rate, and thus we could only identify the new region
as a potential match if we use a $2e$ error rate when categorizing regions. 

\section{Prior Tools Compatability with GATK}\label{suppsec:gatkcompat}

\noindent\textbf{UCSC LiftOver.} UCSC LiftOver only generates Browser Extensible Data
(BED) files when remapping read sets. BED files are incompatible with variant
calling tools, as they lose a lot of information required for variant callers. 

\noindent\textbf{CrossMap.} CrossMap incorrectly handles supplemental alignments,
resulting in duplicate mappings in the BAM file that cannot be used for variant
calling. We have made a few updates to CrossMap as can be found in the forked
repository \url{https://github.com/canfirtina/CrossMap}.

\section{AirLift Index Study} 

We first analyze the AirLift Indices to determine the breakdown of regions
across both old and new references. Supplementary
Table~\ref{tab:region_breakdown}~shows the region sizes (i.e., constant, updated,
retired) that an old reference is comprised of when preprocessed with another
reference. We note that the closer the version numbers between the pair of
references are to each other, 1) the larger the constant region is, and 2) the
smaller the updated region is. This is intuitive as each reference genome
version releases incremental changes to update missing and inaccurate
sequences, so the similarity between consecutive releases would likely be
higher than that between releases further apart. We also observe, as expected,
that the percentage of reads that map to a region in the reference is
correlated with the region size (i.e., larger regions have more reads mapped to
that region). 

Since the most expensive method for remapping in AirLift (i.e., full mapping
via BWA-MEM) is employed \emph{only} for reads that mapped to updated regions
of the old reference or never mapped at all, we can expect a significant
reduction in the mapping time, based on the small updated regions of
Supplementary Table~\ref{tab:region_breakdown}. 

\section{Running AirLift in Multithreaded Mode}\label{suppsec:multithreaded}

If a user specifies multiple threads to execute AirLift, all
multithreaded-enabled tools (i.e., SAMtools, BWA-MEM, and FastRemap, our in-house C++ implementation of CrossMap) used in the execution
pipeline are executed with the specified number of threads. First, AirLift
extracts alignments to the constant regions of the old reference in parallel
using \emph{SAMtools view} with the multithreading option (i.e., “--threads”)
enabled to copy these alignments to a temporary BAM file called
\emph{fastremap\_before.bam} file. Second, AirLift uses \emph{SAMtools index} to
generate the index file for \emph{fastremap\_before.bam} in parallel (using the
“-@” option). Third, AirLift uses FastRemap to update the alignment positions in
the \emph{fastremap\_before.bam} file according to the new reference genome. FastRemap uses the Seqan library, which utilizes as many CPU cores available when executing its function calls. As
FastRemap generates the alignments with the updated positions, the alignments
are piped for sorting using \emph{SAMtools sort}, which runs in parallel (using
the “--threads” option). Fourth, AirLift aligns reads that fall either in the
retired or updated regions of the old reference with a read mapper (e.g.,
BWA-MEM) in multi-threaded mode.  Fifth, AirLift uses \emph{SAMtools merge} to
combine all intermediate alignment results (i.e., mapped and remapped
alignments) in parallel (using the “--threads” option).

\clearpage 

\begin{center}
    \LARGE{Supplementary Tables for\\
AirLift: A Fast and Comprehensive Technique \\for Translating Alignments between Reference Genomes}
\end{center}

\setcounter{table}{0}

 \begin{table}[h!]
      \caption{Annotations in the new reference not covered by reads when remapping reads across reference genomes with a remapping tool that solely relies on chain files (e.g., UCSC LiftOver). }
      \label{tab:annotations}
    \begin{center}
    \begin{tabular}{cccccccc} 
        \footnotesize 
    \setlength{\tabcolsep}{3pt} %
    \renewcommand{\arraystretch}{1.1} %
 	 & & \multicolumn{6}{c}{New Reference}\\
     & & \multicolumn{6}{c}{\textbf{hg19}} \\ 
     \cline{3-8}
     \multirow{4}{*}{\rotatebox[origin=c]{90}{Old Ref.}}&
      \multicolumn{1}{c}{} & gene & exon & stop codon & CDS & start codon & transcript  \\
    \cline{3-8}
     & \multicolumn{1}{c}{\textbf{hg16}} & 3.07&0.92&0.79&0.77&0.72&2.92 \\
     & \multicolumn{1}{c}{\textbf{hg17}} & 1.45&0.36&0.23&0.24&0.24&1.22 \\
     & \multicolumn{1}{c}{\textbf{hg18}} & 0.84&0.12&0.07&0.10&0.10&0.78 \\
     \cline{3-8}
    \multicolumn{8}{c}{} \\ 
     & & \multicolumn{6}{c}{\textbf{hg38}} \\ 
     \cline{3-8}
     \multirow{5}{*}{\rotatebox[origin=c]{90}{Old Ref.}}&
      \multicolumn{1}{c}{} & gene & exon & stop codon & CDS & start codon & transcript  \\
    \cline{3-8}
     & \multicolumn{1}{c}{\textbf{hg16}} & 7.06&2.41&2.13&2.16&2.07&7.03  \\
     & \multicolumn{1}{c}{\textbf{hg17}} & 5.38&1.18&0.93&0.93&0.89&5.13  \\
     & \multicolumn{1}{c}{\textbf{hg18}} & 4.95&0.96&0.73&0.75&0.72&4.72  \\
     & \multicolumn{1}{c}{\textbf{hg19}} & 4.47 &0.74&0.50&0.59&0.53&4.24  \\ 
     \cline{3-8}

     \end{tabular}
     \end{center}
      \label{tab:crossmap_coverage123}

      \begin{minipage}{\textwidth}
      \footnotesize
 Between each pair of reference genomes, we indicate the exact values of specific annotation types (e.g., gene, exon, stop codon, CDS, start codon, transcript) that are ``lost'' when using UCSC LiftOver~\cite{ucscliftover} on a read data set from an old reference (rows) to a new reference (columns). Briefly, 3.07\% of the gene model coordinates in hg16 assembly are not found in hg19, where the loss rate of genes reaches 4.47\% between the most recent two assembly versions (hg19 and hg38).
     \end{minipage}
     
 \end{table}

\begin{table}[h!]
\footnotesize 
\caption{Details of the reference genomes that we use in our evaluations. }
\setlength{\tabcolsep}{1em}
\begin{center}
\label{tab:references}
\begin{tabular}{rcrrc}
\hline
\textbf{Species} & \textbf{Version}  & \textbf{Total \# of Bases} & \textbf{non-N Bases} & \textbf{Release Date} \\\hhline{=====}
\multirow{5}{*}{Human}   & hg16    &  3,091,959,510 & 2,865,086,288 & 2004-02-04 \\
   & hg17    &  3,091,360,260 & 2,865,812,574 & 2004-08-24 \\
   & hg18    &  3,104,054,490 & 2,881,568,385 & 2006-03-03 \\
   & hg19    &  3,137,144,693 & 2,897,293,955 & 2009-02-27 \\
   & hg38    &  3,209,286,105 & 3,049,316,098 & 2013-12-24 \\
\hline
\multirow{5}{*}{C. elegans} & ce2  & 100,291,769 & 100,291,761 & 2004-03-01 \\
 & ce4  & 100,281,244 & 100,281,244 & 2007-01-01 \\ 
 & ce6  & 100,281,426 & 100,281,244 & 2008-05-01 \\
 & ce10 & 100,286,070 & 100,286,070 & 2012-04-13 \\
 & ce11 & 100,286,401 & 100,286,401 & 2013-02-07 \\
\hline
\multirow{3}{*}{Yeast} & sacCer1 & 12,156,302 & 12,156,302 & 2001-10-01 \\
 & sacCer2 & 12,162,995 & 12,162,995 & 2008-06-01 \\
 & sacCer3 & 12,157,105 & 12,157,105 & 2014-12-17 \\
\hline
\end{tabular}
\end{center}
\end{table}

\begin{table}[h!] 
\begin{center}
\caption{\small Read data sets that we use in our evaluations.}
\label{tab:read_data_sets}
\begin{tabular}{rlr}
\hline
\multicolumn{1}{r}{\textbf{Read Data Set}} & \multicolumn{1}{r}{\textbf{Accession No.}}  & \multicolumn{1}{c}{\textbf{Details}}  \\\hhline{===}
Human NA12878 - Illumina  & ERR194147 & 795,505,905 paired-end reads (101bps each, ~50$\times$ coverage) \\
Human NA12878 - Illumina  & ERR262997 & 643,097,275 paired-end reads (101bps each, ~40$\times$ coverage) \\
\hline 
C. elegans N2 - Illumina & SRR3536210 & 78,696,056 paired-end reads (101bps each, 150$\times$ coverage) \\
\hline
Yeast S288C - Illumina & ERR1938683 & 3,318,467 paired-end reads (150bps each, ~82$\times$ coverage) \\
\hline
\end{tabular}
\end{center}
\end{table}

 \begin{table}[!ht]
    \scriptsize 
    \caption{AirLift Execution Time Breakdown} 
     \label{tab:executiontime_breakdown}
    \begin{center}
    
    \begin{tabular}{ccccrrrr}
        \hline
        \multirow{2}{*}{Species} & \multicolumn{2}{c}{Remapping reads} & & \multirow{2}{*}{Preprocessing Time (s)} & \multirow{2}{*}{\textbf{Processing Time (s)}} & \multirow{2}{*}{Postprocessing Time (s)} & \multirow{2}{*}{Total time (s)}\\ \cline{2-3} 
         & From & To & &  \\ \hline
		 \multirow{4}{*}{Human} & hg16 & \multirow{4}{*}{hg38} & & 616.77 (0.47\%) & \textbf{106443.50 (80.90\%)} & 24521.73 (18.64\%) & 131582.01 \\ 
		                                              & hg17 & & & 466.83 (0.37\%) & \textbf{99572.96 (79.94\%)}  & 24519.46 (19.68\%) & 124559.25 \\ 
		                                              & hg18 & & & 505.38 (0.38\%) & \textbf{106366.88 (80.67\%)} & 24980.69 (18.95\%) & 131852.95 \\ 
		                                              & hg19 & & & 437.68 (0.34\%) & \textbf{102745.61 (80.29\%)} & 24785.56 (19.37\%) & 127968.85 \\ \hline 

         \multirow{5}{*}{C. elegans} & ce2 & ce4 & & 22.14 (0.46\%) & \textbf{3452.36 (72.72\%)} & 1272.76 (26.81\%) & 4747.26 \\ \cline{2-8} 
                   & ce4 & \multirow{2}{*}{ce10} & & 1.91 (0.04\%)  & \textbf{3598.83 (75.98\%)} & 1135.70 (23.98\%) & 4736.44 \\ 
                                         & ce6 & & & 3.70 (0.08\%)  & \textbf{3506.43 (73.41\%)} & 1266.10 (26.51\%) & 4776.23 \\ \cline{2-8} 
                   & ce6 & \multirow{2}{*}{ce11} & & 3.08 (0.06\%)  & \textbf{3515.23 (73.38\%)} & 1272.30 (26.56\%) & 4790.61 \\ 
                       & \multirow{1}{*}{ce10} & & & 25.54 (0.52\%) & \textbf{3601.79 (73.55\%)} & 1269.83 (25.93\%) & 4897.16 \\ \hline 

        \multirow{3}{*}{Yeast} & sacCer1 & sacCer2 & & 1.55  (0.80\%) & \textbf{179.74 (93.19\%)} & 11.59\ \ (6.01\%) & 192.88 \\ \cline{2-8} 
              & sacCer1 & \multirow{2}{*}{sacCer3} & & 11.26 (5.41\%) & \textbf{184.35 (88.54\%)} & 12.59\ \ (6.05\%) & 208.20 \\ 
                                       & sacCer2 & & & 8.83  (3.36\%) & \textbf{181.51 (69.09\%)} & 72.37 (27.55\%)   & 262.71 \\ \hline
     \end{tabular}

   \end{center}
   \begin{minipage}{\textwidth}
   \footnotesize
We show for our selected species' reference genomes, human (large), C. elegans (medium), yeast (small) the execution time across pairs of references broken down into preprocessing, processing, and postprocessing times. The execution time is shown for different version pairs of each reference genome (row). The execution times for each stage is measured in seconds and the percentage of the total execution time is shown in parenthesis.
\end{minipage}
 \end{table}

\begin{table}[!ht]
    \footnotesize 
    \caption{AirLift Index Peak Memory Usage} 
     \label{tab:airliftindex_peakmemusage}
    \begin{center}
    \begin{tabular}{cccc}
        \hline
        \multirow{2}{*}{Species} & \multicolumn{2}{c}{Remapping a read set} & \multirow{2}{*}{Peak Memory (MB)} \\ \cline{2-3}
         & From & To & \\ \hline
		 \multirow{4}{*}{Human} & hg16 & \multirow{4}{*}{hg38} & 4,647  \\ 
                              & hg17 &                       & 4,641   \\ 
                              & hg18 &                       & 4,639  \\ 
                              & hg19 &                       & 4,663  \\ \hline 
         \multirow{5}{*}{C. elegans} & ce2 & ce4 & 101 \\ \cline{2-4} 
                                     & ce4 & \multirow{2}{*}{ce10} & 150  \\ 
                                     & ce6 &                       & 150  \\ \cline{2-4} 
                                     & ce6 & \multirow{2}{*}{ce11} & 151  \\ 
                                     & ce10 &                      & 150  \\ \hline 
        \multirow{3}{*}{Yeast} & sacCer1 & sacCer2 & 20 \\ \cline{2-4} 
                               & sacCer1 & \multirow{2}{*}{sacCer3} & 21 \\ 
                               & sacCer2 &                          & 21 \\ \hline
     \end{tabular}
   \end{center}
   \begin{minipage}{\textwidth}
   \footnotesize
   We show for our selected species' reference genomes, human (large), C. elegans (medium), yeast (small) the peak memory usage across pairs of references during the preprocessing (AirLift Index construction) step.
\end{minipage}
 \end{table}

 \begin{table}[!ht]
    \footnotesize 
    \caption{Breakdown of Region Labels for Each Pair of Reference Genomes.} 
     \label{tab:region_breakdown}
    \begin{center}

    \begin{tabular}{cccccc}
        \hline
        \multirow{2}{*}{Species} & \multicolumn{2}{c}{Remapping a read set} & \multirow{2}{*}{Constant (\%)} & \multirow{2}{*}{Updated (\%)} & \multirow{2}{*}{Retired (\%)} \\ \cline{2-3}
         & From & To & & & \\ \hline
		 \multirow{4}{*}{Human} & hg16 & \multirow{4}{*}{hg38} & 85.7475 & 14.1867 & 0.0659 \\ 
		 & hg17 & & 86.5513 & 13.4106 & 0.0382  \\ 
		 & hg18 & & 86.6874 & 13.2485 & 0.0641  \\ 
		 & hg19 & & 87.1995 & 12.7344 & 0.0660  \\ \hline 
         \multirow{5}{*}{C. elegans} & ce2 & ce4 & 99.9862 & 0.0109 & 0.0028 \\ \cline{2-6} 
         & ce4 & \multirow{2}{*}{ce10} & 99.9738 & 0.0222 & 0.0040 \\ 
         & ce6 & & 99.9770 & 0.0191 & 0.0040  \\ \cline{2-6} 
         & ce6 & \multirow{2}{*}{ce11} & 99.8262 & 0.1587 & 0.0151 \\ 
         & \multirow{1}{*}{ce10} & & 99.8505 & 0.1428 & 0.0067  \\ \hline 
        \multirow{3}{*}{Yeast} & sacCer1 & sacCer2 & 90.2503 & 8.7276 & 1.0220  \\ \cline{2-6} 
         & sacCer1 & \multirow{2}{*}{sacCer3} & 99.4449 & 0.5297 & 0.0254  \\ 
         & sacCer2 & & 99.5459 & 0.4289 & 0.0252 \\ \hline
     \end{tabular}

   \end{center}
   \begin{minipage}{\textwidth}
   \footnotesize
We show for our selected species' reference genomes, human (large), C. elegans (medium), yeast (small) how versions of the reference genome (row) are comprised of distinct regions (i.e., constant, updated, retired) in relation to a more recent version of the species. Each cell contains the percentage of the old reference genome that each category of regions (columns) comprises. 
\end{minipage}
 \end{table}

 \begin{table}[!ht]
    \scriptsize
    \centering
    \caption{Execution Time Breakdown for Remapping a Read Set by Case} 
     \label{tab:executiontime_breakdown_step}
    \begin{center}
    \begin{tabular}{ccccrrrr}
        \hline
        \multirow{2}{*}{Species} & \multicolumn{2}{c}{Remapping reads} & & \multicolumn{4}{c}{Time to remap reads that originally mapped to region (s):} \\ \cline{2-3} \cline{5-8}
         & From & To & & 1. Constant & 2. Updated & 3. Retired & \textbf{Total} \\ \hline
		 \multirow{4}{*}{Human} & hg16 & \multirow{4}{*}{hg38} & & 88524.04 (83.17\%) & 17457.64 (16.40\%) & 461.83 (0.43\%) & \textbf{106443.51} \\ 
		                                              & hg17 & & & 88832.96 (89.21\%) & 10416.94 (10.46\%) & 323.06 (0.32\%) & \textbf{99572.96} \\ 
		                                              & hg18 & & & 88963.14 (83.64\%) & 16926.51 (15.91\%) & 477.23 (0.45\%) & \textbf{106366.88} \\ 
		                                              & hg19 & & & 87779.77 (85.43\%) & 14539.23 (14.15\%) & 426.61 (0.42\%) & \textbf{102745.61} \\ \hline 

         \multirow{5}{*}{C. elegans} & ce2 & ce4 & & 3435.98 (99.53\%) & 16.37 (0.47\%)  & 0.00 (0.00\%) & \textbf{3452.36} \\ \cline{2-8} 
                   & ce4 & \multirow{2}{*}{ce10} & & 3461.82 (96.19\%) & 137.00 (3.81\%) & 0.00 (0.00\%) & \textbf{3598.83} \\ 
                                         & ce6 & & & 3504.06 (99.93\%) & 2.36 (0.07\%)   & 0.00 (0.00\%) & \textbf{3506.43} \\ \cline{2-8} 
                   & ce6 & \multirow{2}{*}{ce11} & & 3513.08 (99.94\%) & 2.10 (0.06\%)   & 0.00 (0.00\%) & \textbf{3515.19} \\ 
                       & \multirow{1}{*}{ce10} & & & 3581.78 (99.44\%) & 19.97 (0.55\%)  & 0.00 (0.00\%) & \textbf{3601.79} \\ \hline 

        \multirow{3}{*}{Yeast} & sacCer1 & sacCer2 & & 179.17 (99.68\%) & 0.56 (0.31\%) & 0.00 (0.00\%) & \textbf{179.74} \\ \cline{2-8} 
              & sacCer1 & \multirow{2}{*}{sacCer3} & & 181.11 (98.24\%) & 3.21 (1.74\%) & 0.00 (0.00\%) & \textbf{184.35} \\ 
                                       & sacCer2 & & & 178.61 (98.40\%) & 2.87 (1.58\%) & 0.00 (0.00\%) & \textbf{181.51} \\ \hline
     \end{tabular}
   \end{center}
   \begin{minipage}{\textwidth}
   \footnotesize
We show for our selected species' reference genomes, human (large), C. elegans (medium), yeast (small) the execution time breakdown of remapping a read set. The execution time is shown for different versions of each reference genome (row) and is shown for the four remapping cases: 1) reads that originally mapped to a constant region, 2) reads that originally mapped to an updated region, 3) reads that originally mapped to a retired region, and 4) reads that never mapped to the old reference (i.e., unmapped). The remap time for each case is measured in seconds and the percentage of the full remapping time is shown in parenthesis. 
\end{minipage}
 \end{table}

 \begin{table}[!ht]
    \scriptsize 
    \caption{Number of Reads Originally Mapped within Each AirLift Case} 
     \label{tab:reads_breakdown_step}
    \begin{center}
    \begin{tabular}{ccccrrrrr}
        \hline
        \multirow{2}{*}{Species} & \multicolumn{2}{c}{Remapping reads} & & \multicolumn{5}{c}{Number of reads that originally mapped to region:} \\ \cline{2-3} \cline{5-9}
         & From & To & & 1. Constant & 2. Updated & 3. Retired & 4. Unmapped & \textbf{Total} \\ \hline
		 \multirow{4}{*}{Human} & hg16 & \multirow{4}{*}{hg38} & & 2866050090 (86.57\%) & 401163904 (12.12\%) & 408920 (0.0124\%)  & 43205312 (1.30\%) & \textbf{3310828226} \\ 
		                                              & hg17 & & & 2869062015 (86.78\%) & 396473981 (11.99\%) & 126607 (0.0038\%) & 40431845 (1.22\%) & \textbf{3306094448} \\ 
		                                              & hg18 & & & 2870008151 (88.04\%) & 350216312 (10.74\%) & 126040 (0.0039\%) & 39551799 (1.21\%) & \textbf{3259902302} \\ 
		                                              & hg19 & & & 2870919584 (90.88\%) & 249423596 (7.90\%)  & 60300 (0.0019\%)  & 38781755 (1.23\%) & \textbf{3159185235} \\ \hline 

         \multirow{5}{*}{C. elegans} & ce2 & ce4 & & 155088879 (98.43\%) & 59432 (0.038\%) & 0 (0.000\%) & 2413178 (1.53\%) & \textbf{157561489} \\ \cline{2-9} 
                   & ce4 & \multirow{2}{*}{ce10} & & 155091089 (98.41\%) & 91980 (0.058\%) & 0 (0.000\%) & 2413157 (1.53\%) & \textbf{157596226} \\ 
                                         & ce6 & & & 155091089 (98.46\%) & 9444 (0.006\%)  & 0 (0.000\%) & 2413157 (1.03\%) & \textbf{157513690} \\ \cline{2-9} 
                   & ce6 & \multirow{2}{*}{ce11} & & 155091090 (98.47\%) & 1040 (0.001\%)  & 0 (0.000\%) & 2413156 (1.53\%) & \textbf{157505286} \\ 
                       & \multirow{1}{*}{ce10} & & & 155094686 (98.47\%) & 0 (0.000\%)     & 0 (0.000\%) & 2409560 (1.53\%) & \textbf{157504246} \\ \hline 

        \multirow{3}{*}{Yeast} & sacCer1 & sacCer2 & & 6509230 (97.79\%) & 17676 (0.27\%) & 0 (0.00\%) & 129581 (1.95\%) & \textbf{6656487} \\ \cline{2-9} 
              & sacCer1 & \multirow{2}{*}{sacCer3} & & 6509230 (97.36\%) & 63404 (0.95\%) & 0 (0.00\%) & 129581 (1.93\%) & \textbf{6702215} \\ 
                                       & sacCer2 & & & 6509230 (97.79\%) & 46652 (0.70\%) & 0 (0.00\%) & 129576 (1.94\%) & \textbf{6685458} \\ \hline
     \end{tabular}
   \end{center}
   \begin{minipage}{\textwidth}
   \footnotesize
We show for our selected species' reference genomes, human (large), C. elegans (medium), yeast (small) the number of reads in a read set that is mapped by each AirLift case. The number of reads is shown for different versions of each reference genome (row) and is shown for four remapping cases: 1) reads that originally mapped to a constant region, 2) reads that originally mapped to an updated region, 3) reads that originally mapped to a retired region, and 4)~reads that were originally unmapped. The percentage of the full read set is shown in parentheses. 
\end{minipage}
\end{table}

\begin{table}[!ht]
\centering
\caption{Ratio of Reads Remapped from Each AirLift Case} 

    \begin{tabular}{ccccrrrr}
        \hline
        \multirow{2}{*}{Species} & \multicolumn{2}{c}{Remapping reads} & & \multicolumn{4}{c}{Ratio of reads that remapped from region:} \\ \cline{2-3} \cline{5-8}
         & From & To & & 1. Constant & 2. Updated & 3. Retired & 4. Unmapped \\ \hline
		 \multirow{4}{*}{Human} & hg16 & \multirow{4}{*}{hg38} & & 1.0 & 0.999 & 0.0195 & 0.00221\\ 
		                                              & hg17 & & & 1.0 & 0.999 & 0.0225 & 0.00153\\ 
		                                              & hg18 & & & 1.0 & 0.998 & 0.017 & 0.00101\\ 
		                                              & hg19 & & & 1.0 & 0.999 & 0.0209 & 0.00002\\ \hline 

         \multirow{5}{*}{C. elegans} & ce2 & ce4 & & 1.0 & 0.998 & NA & 0.00001\\\cline{2-8} 
                   & ce4 & \multirow{2}{*}{ce10} & & 1.0 & 0.9975 & NA & 0.00119\\ 
                                         & ce6 & & & 1.0 & 0.999 & NA & 0.00119\\\cline{2-8} 
                   & ce6 & \multirow{2}{*}{ce11} & & 1.0 & 0.997 & NA & 0.00123\\ 
                       & \multirow{1}{*}{ce10} & & & 1.0 & NA & NA & 0.00004 \\\hline 

        \multirow{3}{*}{Yeast} & sacCer1 & sacCer2 & & 1.0 & 0.9959 & NA & 0.00008\\\cline{2-8} 
              & sacCer1 & \multirow{2}{*}{sacCer3} & & 1.0 & 0.9976 & NA & 0.00138 \\
                                       & sacCer2 & & & 1.0 & 0.9981 & NA & 0.00130\\\hline
     \end{tabular}
   \begin{minipage}{\textwidth}
   \footnotesize
We show for our selected species' reference genomes, human (large), C. elegans (medium), yeast (small) the ratio of reads within a particular AirLift case that is remapped to the new reference genome. The ratio of reads is shown for different versions of each reference genome (row) and is shown for four remapping cases: 1) reads that originally mapped to a constant region, 2) reads that originally mapped to an updated region, 3) reads that originally mapped to a retired region, and 4)~reads that were originally unmapped.
\end{minipage}

\label{tab:reads_breakdown_mapped_step}
\end{table}

 \begin{table}[!ht]
    \scriptsize 
    \caption{Execution Time per Read when Remapping a Read Set by Case} 
     \label{tab:executiontime_breakdown_region}
    \begin{center}
    
    \begin{tabular}{ccccrrrr}
        \hline
        \multirow{2}{*}{Species} & \multicolumn{2}{c}{Remapping reads} & & \multicolumn{4}{c}{Avg time to remap a read originally mapped to region (us)} \\ \cline{2-3} \cline{5-8}
         & From & To & & 1. Constant & 2. Updated & 3. Retired & 4. Unmapped \\ \hline
		 \multirow{4}{*}{Human} & hg16 & \multirow{4}{*}{hg38} & & 30.887 & 43.517 & 1129.390 & 225.320  \\ 
		                                              & hg17 & & & 30.962 & 26.274 & 2551.676 & 195.822  \\ 
		                                              & hg18 & & & 30.998 & 48.332 & 3786.338 & 185.492  \\ 
		                                              & hg19 & & & 30.575 & 58.291 & 7074.793 & 175.526  \\ \hline 

         \multirow{5}{*}{C. elegans} & ce2 & ce4 & & 22.155 & 275.508  & 5.882 & 20.989  \\ \cline{2-8} 
                   & ce4 & \multirow{2}{*}{ce10} & & 22.321 & 1489.454 & -     & 22.435  \\ 
                                         & ce6 & & & 22.594 & 249.894  & -     & 22.435  \\ \cline{2-8} 
                   & ce6 & \multirow{2}{*}{ce11} & & 22.652 & 2019.231 & -     & 21.507  \\ 
                       & \multirow{1}{*}{ce10} & & & 23.094 & -        & -     & 21.369  \\ \hline 

        \multirow{3}{*}{Yeast} & sacCer1 & sacCer2 & & 27.526 & 31.681 & - & 10.958 \\ \cline{2-8} 
              & sacCer1 & \multirow{2}{*}{sacCer3} & & 27.824 & 50.628 & - & 10.958 \\ 
                                       & sacCer2 & & & 27.439 & 61.519 & - & 10.959 \\ \hline
     \end{tabular}

   \end{center}
   \begin{minipage}{\textwidth}
   \footnotesize
We show for our selected species' reference genomes, human (large), C. elegans (medium), yeast (small) the average execution time to remap a single read depending on the AirLift case. The execution time is shown for different versions of each reference genome (row) and is shown for the four remapping cases: 1) reads that originally mapped to a constant region, 2) reads that originally mapped to an updated region, 3) reads that originally mapped to a retired region, and 4) reads that never mapped to the old reference (i.e., unmapped). The average remap time for each read is measured in microseconds. Each value is calculated by dividing the corresponding cell in Table S6 by the corresponding cell in Table S7. 
\end{minipage}
 \end{table}

\clearpage

\begin{center}
    \LARGE{Supplementary Figures for\\
AirLift: A Fast and Comprehensive Technique \\for Translating Alignments between Reference Genomes}
\end{center}

 \begin{figure}[h]
    \centering
    \captionsetup{type=figure} 
    \includegraphics[width=0.6\linewidth]{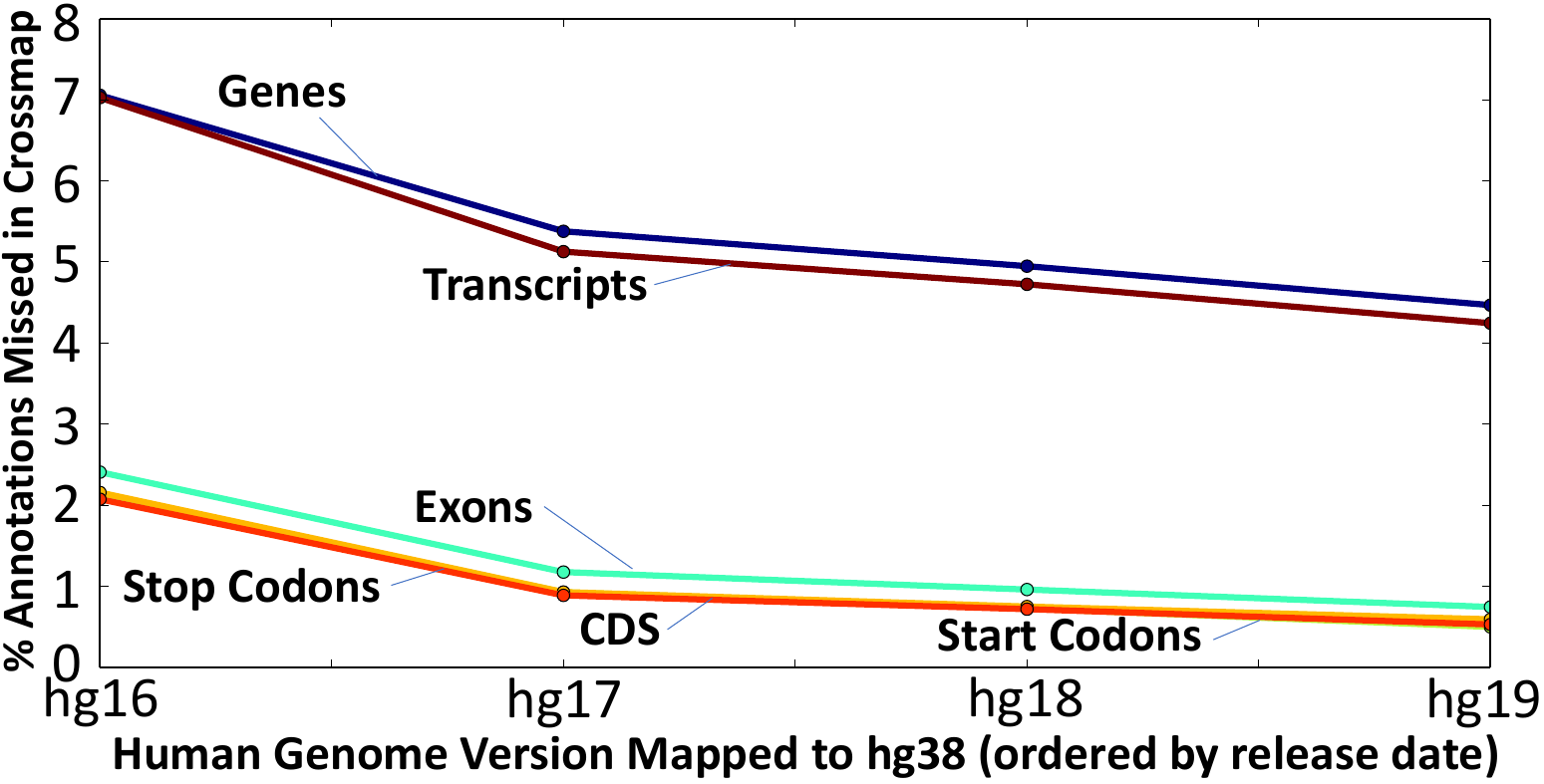}
    \caption{Percentage of different annotations missed when remapping reads from an old reference (x-axis) to the latest reference (hg38), using a remapping tool that solely relies on existing chain files (e.g., UCSC LiftOver~\cite{ucscliftover})}.
    \label{fig:crossmap_analysis1}
\end{figure}

\vspace{2cm}
\begin{figure}[htb]
    \centering
    \captionsetup{type=figure} 
    \includegraphics[width=0.8\linewidth]{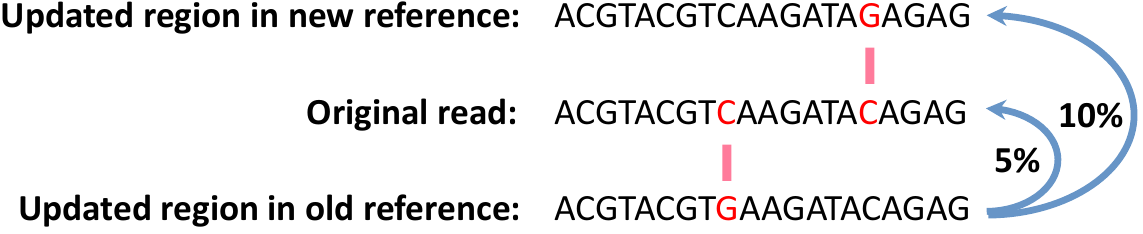}
    \caption{In order to comprehensively account for possible mappings of a read that previously mapped to an old reference genome, we create a lookup table describing the similarity between two reference genomes, using 2$\times$ the alignment error acceptance rate. As an example, if a read aligns to a location in the old reference genome with a 5\% error rate (1 substitution in 20 base pairs), it is possible for the same read to map to a location in the new reference genome (with a 5\% error rate) whose sequence is 10\% different (2 substitutions in 20 base pairs) from the sequence in the old reference genome.}
    \label{fig:remap_2e}
\end{figure}

\clearpage

\bibliographystylesupp{IEEEtran}
\bibliographysupp{main}

\end{document}